\documentclass[twocolumn,aps,prc,showpacs,floatfix,superscriptaddress]{revtex4}
\usepackage{graphicx}
\usepackage{amsmath}

\pdfoutput=1

\newcommand{\Cerenkov}{$\breve{\text{C}}$erenkov }

\newcommand{\LfourHe}{\mbox{$_{\Lambda}^4\text{He}$ }}
\newcommand{\LfiveHe}{\mbox{$_{\Lambda}^5\text{He}$ }}
\newcommand{\LfourH}{\mbox{$_{\Lambda}^4\text{H}$ }}
\newcommand{\fourHe}{\mbox{$^4$He }}
\newcommand{\Km}{\mbox{$K^-$ }}
\newcommand{\pim}{\mbox{$\pi^-$ }}
\newcommand{\piz}{\mbox{$\pi^{\circ}$ }}
\newcommand{\Kpi}{\mbox{$K\pi$ }}

\newcommand{\LN}{\mbox{$\Lambda N$ }}
\newcommand{\LNN}{\mbox{$\Lambda N \negthinspace N$ }}
\newcommand{\Lp}{\mbox{$\Lambda p$ }}
\newcommand{\Ln}{\mbox{$\Lambda n$ }}
\newcommand{\Lpp}{\mbox{$\Lambda p p$ }}

\newcommand{\Kmnp}{\mbox{$K^- \rightarrow \mu^- \bar{\nu}_\mu \pi^\circ$ }}

\newcommand{\delIeq}{\mbox{$\Delta I = \frac{1}{2}$ }}
\newcommand{\delIeqthree}{\mbox{$\Delta I = \frac{3}{2}$ }}

\newcommand{\Gpim}{\mbox{$\Gamma_{\pi^-}$ }}
\newcommand{\Gnpr}{\mbox{$\Gamma_n / \Gamma_p$ }}

\begin{document}

\title{Weak decays of $_\Lambda^{4}{\rm He}$}

\author{J.D. Parker}
\altaffiliation[Present address: ]
 {Department of Physics, Kyoto University, Kyoto 606-8502, Japan.}
\email[]{jparker@scphys.kyoto-u.ac.jp}
\author{M.J. Athanas}
\altaffiliation[Present address: ]
 {VAST Scientific, Cambridge, MA 02139, USA}
\author{P.D. Barnes}
\altaffiliation[Present address: ]
 {Los Alamos National Laboratory, Los Alamos, NM 87544, USA}
\affiliation{Carnegie Mellon University, Pittsburgh, PA 15213, USA}
\author{S. Bart}
\affiliation{Brookhaven National Laboratory, Upton, NY 11973, USA}
\author{B. Bassalleck}
\affiliation{University of New Mexico, Albuquerque, NM 87131, USA}
\author{R. Chrien}
\affiliation{Brookhaven National Laboratory, Upton, NY 11973, USA}
\author{G. Diebold}
\author{G.B. Franklin}
\affiliation{Carnegie Mellon University, Pittsburgh, PA 15213, USA}
\author{K. Johnston}
\altaffiliation[Present address: ]
 {Louisiana Tech University, Ruston, LA 71272, USA}
\affiliation{University of Houston, Houston, TX 77004, USA}
\author{P. Pile}
\affiliation{Brookhaven National Laboratory, Upton, NY 11973, USA}
\author{B. Quinn}
\author{F. Rozon}
\affiliation{Carnegie Mellon University, Pittsburgh, PA 15213, USA}
\author{R. Sawafta}
\altaffiliation[Present address: ]
 {Quartek Corporation, Greensboro, NC 27410, USA}
\affiliation{Brookhaven National Laboratory, Upton, NY 11973, USA}
\author{R.A. Schumacher}
\affiliation{Carnegie Mellon University, Pittsburgh, PA 15213, USA}
\author{R. Stearns}
\affiliation{Vassar College, Poughkeepsie, NY 12604, USA}
\author{I. Sukaton}
\affiliation{Carnegie Mellon University, Pittsburgh, PA 15213, USA}
\author{R. Sutter}
\affiliation{Brookhaven National Laboratory, Upton, NY 11973, USA}
\author{J.J. Szymanski}
\altaffiliation[Present address: ]
 {Los Alamos National Laboratory, Los Alamos, NM 87544, USA}
\affiliation{Indiana University Cyclotron Facility, Bloomington, IN 47408, USA}
\author{V.J. Zeps}
\altaffiliation[Present address: ]
 {Bluegrass Community and Technical College, Lexington, KY 40506, USA}
\affiliation{Carnegie Mellon University, Pittsburgh, PA 15213, USA}

\date[Revised ]{September 13, 2007}

\begin{abstract}
We measured the
lifetime and the mesonic and nonmesonic decay rates of the \LfourHe 
hypernucleus.  
The hypernuclei were created using a 750~MeV/c momentum \Km beam on a liquid
\fourHe target by the reaction 
$^4\text{He}(K^-,\pi^-)^4_\Lambda\text{He}$.
The \LfourHe lifetime was directly measured using protons from 
$\Lambda p \rightarrow n p$ nonmesonic decay (also referred to as 
proton-stimulated decay) and
was found to have a value of $\tau = 245 \pm 24$~ps.
The mesonic decay rates were determined from the observed numbers of $\pi^-$'s
and $\pi^\circ$'s as $\Gamma_{\pi^-}/\Gamma_{tot} = 0.270 \pm 0.024$ and
$\Gamma_{\pi^\circ}/\Gamma_{tot} = 0.564 \pm 0.036$, respectively,
and the values of the proton- and neutron-stimulated decay rates were 
extracted as
$\Gamma_p/\Gamma_{tot} = 0.169 \pm 0.019$ and 
$\Gamma_n/\Gamma_{tot} \leq 0.032$ (95\% CL), respectively.
The effects of final-state interactions and possible three-body \LNN decay
contributions were studied in the context of a simple model of 
nucleon-stimulated decay.
Nucleon-nucleon coincidence events were observed and 
were used in the determination of the nonmesonic branching fractions.
The implications of the results of this analysis were considered
for the empirical \delIeq rule and the decay rates of the \LfourH hypernucleus.
\end{abstract}


\maketitle

\section{INTRODUCTION}

Studies of the hyperon-nucleon interaction investigate weak processes that
are not well understood in multiple baryon systems.
In particular, the study
of nonmesonic decays of light hypernuclei is the only tractable method of
investigating the $\Delta S=1$ weak baryon-baryon interaction.  By
comprehensive measurements in s-shell hypernuclei, various $\Lambda N$
initial states can be investigated.  
Few-body nuclear structure calculations can
then relate the decay rates to the underlying weak $\Lambda N$ interactions.
These studies are complementary to investigations of the parity-violating
contributions to the nucleon-nucleon interaction.  However, the $\Lambda N$
system can be prepared in a particle-stable state so that only weak processes
contribute to the decay rate.

A particularly interesting aspect of $\Delta S=1$ transitions is the
empirical $\Delta{\rm I}=\frac12$ rule.
We note that if strong interactions are neglected, the $\Delta S=1$
non-leptonic decays of hyperon systems would be expected to be dominated by
a weak quark-level Hamiltonian based on one W-exchange of the form:
\begin{equation}
H_{\rm weak}=\frac{G_f}{\sqrt2}\sin\theta_c\cos\theta_c
             \left(\overline{u}\gamma_\mu(1-\gamma_5)s
             \overline{d}\gamma^\mu(1-\gamma_5)u\right).
\end{equation}
The isospin character of this fundamental Hamiltonian does not favor 
$\Delta{\rm I}=\frac12$ weak transitions over the $\Delta{\rm I}=\frac32$ ones.
However, it has long been recognized that the $\Delta{\rm I}=\frac12$
transitions dominate all observed non-leptonic decays of strange hadrons.
Although it is generally believed that the presence of the strong interaction
accounts for this empirical $\Delta{\rm I}=\frac12$ selection rule,
the various explanations for the dominance of the 
$\Delta{\rm I}=\frac12$ amplitudes~\cite{Okun82,Isgur90} have indicated that
the mechanism may be
specific to the given system and not a universal feature 
of the $\Delta S=1$ weak Hamiltonian. 
It remains an open question whether all $\Delta{\rm S}=1$ non-leptonic 
decays obey the rule or it is specific to the pionic decay channels (the only
tested class of non-leptonic decays).  The study of hypernuclear decays
provides an opportunity to answer this question by providing the only
experimentally accessible non-pionic test of the $\Delta{\rm I}=\frac12$
rule.

The total decay width of a hypernucleus includes contributions from 
mesonic and nonmesonic decay modes.
The mesonic decay modes, defined experimentally by the observation of the
appropriate pion in the final state, are dominated by single-body processes
analogous to the decay of the free $\Lambda$,
$\Lambda \rightarrow \pi^- p$ and $\Lambda \rightarrow \pi^0 n$, along with
additional multi-body contributions such as $\Lambda p \rightarrow nn\pi^+$.
The nonmesonic modes are expected to be dominated by the proton-stimulated
reaction $\Lambda p \rightarrow n p$ and neutron-stimulated reaction
$\Lambda n \rightarrow n n$ with
rates $\Gamma_p$ and $\Gamma_n$, respectively.  Other multi-body interactions
such as $\Lambda NN \rightarrow NNN$ may also contribute with rate $\Gamma_{mb}$. 
Determination of the nucleon-stimulated decay rates is more complicated than
the mesonic case and is ultimately model dependent.
The total decay width can be written in terms of the widths of these
mesonic and nonmesonic decay modes as:
\begin{equation}
\begin{array}{llccc}
\Gamma_{\rm total} & = & \Gamma_{\rm mesonic}&+&\Gamma_{\rm nonmesonic}, \\
                   & = & \overbrace{\Gamma_{\pi^-}+\Gamma_{\pi^0}+
                   \Gamma_{\pi^+}} &+&
                   \overbrace{\Gamma_p+\Gamma_n+\Gamma_{mb}},
\end{array} 
\label{eqn-total}
\end{equation}
where contributions from semileptonic and weak radiative $\Lambda$ decays
have been neglected as they contribute only about 0.3\% to the total free
$\Lambda$ decay width~\cite{PDG06}.

The two-body nonmesonic decay modes, $\Lambda N \rightarrow NN$,
are readily distinguishable from the mesonic decay modes, $\Lambda \rightarrow
\pi N$, because of the large
energy ($M_\Lambda - M_n =176~{\rm MeV}$) available for the final-state
nucleons.
These modes are sensitive to weak interaction couplings 
(such as $g_{\Lambda N \rho}$ or $g_{NNK}$) not available to the free 
hyperon decays. Also, it has been suggested that several of
the weak $\Lambda N \rightarrow NN$ amplitudes are dominated by
direct-quark processes~\cite{Maltman94,Inoue98,Sasaki05} in which no 
intermediate meson is present in the interaction. 
Such direct processes may not adhere to the 
$\Delta I=\frac12$ rule.

\begin{figure}[t]
\includegraphics[width=8.6 cm,clip]{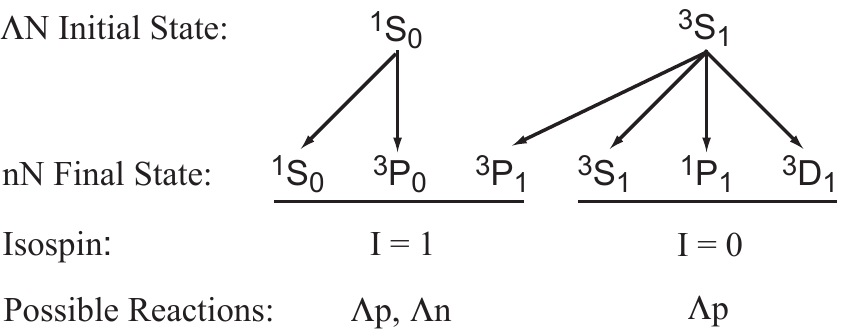}
\caption{Spin-Isospin selection rules for the $\Lambda N \rightarrow nN$
process in 1$s$-shell hypernuclei.
\label{spinisospin}}
\end{figure}

In nuclei the $\Lambda$-particle is not Pauli blocked, thus hypernuclei 
generally decay
with the $\Lambda$-particle initially in the lowest $1s$-state. 
In order to determine properties of the underlying 
$\Lambda N \rightarrow NN$ interaction, it is convenient to 
consider systems that confine the nucleons to the lowest orbitals.
Figure~\ref{spinisospin} shows the spin-isospin
character of the initial ($\Lambda N$) and final ($nN$) states for $s$-shell
hypernuclei in a single-particle shell model picture.  
The I=1
final states are accessible to both proton- and neutron-stimulated decays, 
while the I=0 final states are available only for proton-stimulated decay.
Further complications arise as one increases $A$ because the high 
probability of final-state interactions dilutes the correlation between
nucleons observed and nucleons responsible for the decay.
Table~\ref{tbl-allowed} shows that a complete set of measurements
of the nonmesonic decay widths for the s-shell hypernuclei allows one to
isolate specific $\Lambda N$ initial states.
Thus, 
one can determine the spin-isospin structure of the fundamental
$\Lambda N \rightarrow nN$ weak interaction by combining measurements
of these few-body hypernuclei with detailed finite nucleus calculations that
take into account the differences in the initial and final state phase-space.
However, at this time the experimental data are incomplete and often have
large estimated errors.

\begin{table}
\caption{\label{tbl-allowed} Allowed initial states for $1s-$shell 
hypernuclei.}
\begin{ruledtabular}
\begin{tabular}{ccc}
Species       &   $\Lambda n$ & $\Lambda p$ \\
\hline
$^5_\Lambda{\rm He}$ & $^3S_1,~^1S_0$ & $^3S_1,~^1S_0$ \\
$^4_\Lambda{\rm He}$ & $^1S_0$        & $^3S_1,~^1S_0$ \\
$^4_\Lambda{\rm H}$  & $^3S_1,~^1S_0$ & $^1S_0$        \\
$^3_\Lambda{\rm H}$  & $^3S_1,~^1S_0$ & $^3S_1,~^1S_0$ \\
\end{tabular}
\end{ruledtabular}
\end{table}

A phenomenological model first put forth by 
Block and Dalitz~\cite{Block63,Reinhard92} can be used to
relate the nonmesonic decay rates of the $s$-shell hypernuclei listed in
Table~\ref{tbl-allowed}.
Specifically, the current experimental results for \LfourHe (this work) 
and \LfiveHe (Outa {\em et al}~\cite{Outa05}) can
be extended to the \LfourH hypernucleus, for which measurements are currently
scarce.
The model employs the mean nucleon density, $\rho_A$, at the 
location of the $\Lambda$ and makes several assumptions concerning the \LN 
interaction:
(1) the nucleon-stimulated $\Lambda$ decay is treated as incoherent,
(2) final-state interactions are not included, and
(3) three-body nucleon-stimulated decays are neglected.
These assumptions are seen to be
adequate at the present level of accuracy of the experimental quantities.
Defining the quantity $R_{N \negthinspace S}$ as the rate for an 
$N$-stimulated decay originating from \LN relative spin state $S$, 
the nonmesonic decay rates for the $s$-shell hypernuclei $_\Lambda^4$H,
$_\Lambda^4$He, and $_\Lambda^5$He are derived as:
\begin{eqnarray}
\label{eq:gnm}
\Gamma_{nm}(^4_\Lambda\text{H}) 
 &=& \frac{\rho_4}{6} \left( R_{n0} + 3 R_{n1} + 2 R_{p0} \right), 
\label{eq:h4}\\
\Gamma_{nm}(^4_\Lambda\text{He}) 
 &=& \frac{\rho_4}{6} \left( 2 R_{n0} + R_{p0} + 3 R_{p1} \right), 
\label{eq:he4}\\
\Gamma_{nm}(^5_\Lambda\text{He}) 
 &=& \frac{\rho_5}{8} \left( R_{n0} + 3 R_{n1} + R_{p0} + 3 R_{p1} \right),
\label{eq:he5}
\end{eqnarray}
where the mean nucleon density, $\rho_A$, is defined as:
\begin{equation}
\rho_A \equiv \int \rho_A(\vec{r}) \left|\psi(\vec{r})\right|^2 d\vec{r},
\end{equation}
for the nucleon density, $\rho_A(\vec{r})$, and the $\Lambda$ wave function,
$\psi(\vec{r})$.
Also, it has been assumed that the mean nucleon density, $\rho_4$, has the 
same value for both $^4_\Lambda\text{H}$ and $^4_\Lambda\text{He}$.

Taking the ratio of the rates of neutron-stimulated \LfourHe decay to 
proton-stimulated \LfourH decay yields:
\begin{equation}
\label{eq:hp}
\frac{\Gamma_n(^4_\Lambda\text{He})}{\Gamma_p(^4_\Lambda\text{H})}
 = \frac{R_{n0}}{R_{p0}} = 2,
\end{equation}
where the value of 2 is the \delIeq rule prediction for this ratio.
Similarly, an expression for 
the \LfourH neutron-stimulated rate can be found by considering the ratio
of neutron-stimulated \LfourH decay to proton-stimulated \LfourHe decay:
\begin{equation}
\label{eq:hn}
\frac{\Gamma_n(^4_\Lambda\text{H})}{\Gamma_p(^4_\Lambda\text{He})}
 = \frac{R_{n0} + 3 R_{n1}}{R_{p0} + 3 R_{p1}}
 = \frac{\Gamma_n}{\Gamma_p}(^5_\Lambda\text{He}),
\end{equation}
independent of any assumptions about the \delIeq rule.

In this article, we present the results of BNL experiment E788, which 
measured the lifetime and partial widths of the \LfourHe hypernucleus.
Section~\ref{sec:exp} presents the experimental
apparatus used for these measurements, and a discussion of the
out-of-beam tracking and particle identification algorithm is given in 
Section~\ref{sec:pid}.
The excitation energy spectra for the 
$^4\text{He}(K^-,\pi^-)^4_\Lambda\text{He}$ production reaction 
are discussed in Section~\ref{sec:ee}.
The extraction of the lifetime of $^{4}_\Lambda{\rm He}$ and 
particle-emission spectra are presented in Sections~\ref{sec:life} and 
\ref{sec:ke}, respectively.
The determination of the \LfourHe mesonic and nonmesonic partial decay rates 
is discussed in Section~\ref{sec:rates}.
In the case of the nonmesonic proton- and neutron-stimulated decays,
the rates are found from both the single-particle kinetic 
energy spectra of Section~\ref{sec:ke} as well as from observed 
multiple-nucleon coincidence events, giving two essentially independent 
determinations.
A comparison with the results of other experiments and theoretical calculations
and the extension of these results to the \LfourH hypernucleus is presented in
Section~\ref{sec:conc}.

\section{EXPERIMENTAL SETUP}
\label{sec:exp}

\begin{figure}[ht]
\includegraphics[width=8.6 cm,clip]{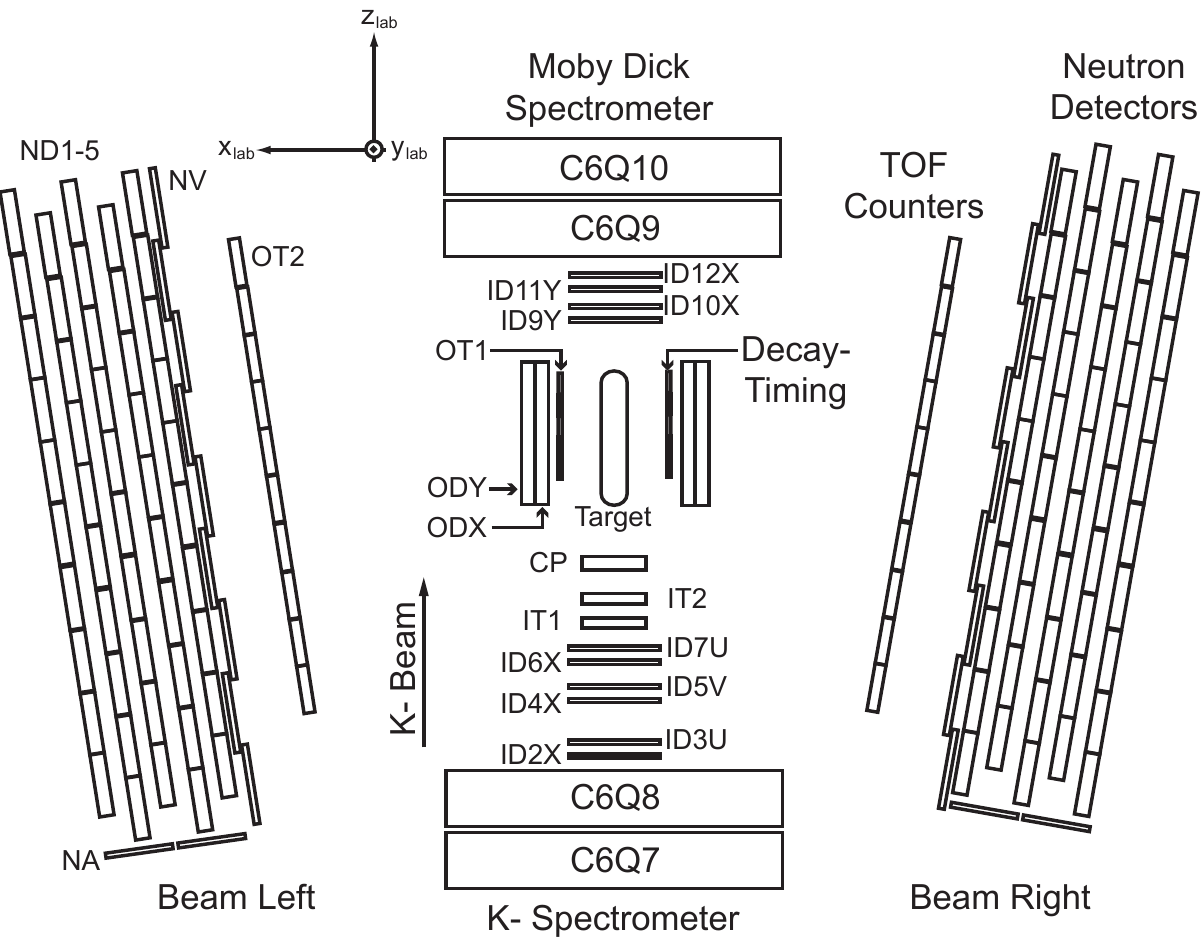}
\caption{
Shown here is the E788 experimental apparatus used to carry out the \LfourHe 
hypernuclear decay measurement.
Also shown are the focusing quadrapoles at the exit of the \Km spectrometer
and at the entrance of the Moby Dick spectrometer.
\label{layout}}
\end{figure}

The partial decay rates and lifetime of $^{4}_\Lambda{\rm He}$ were measured 
using the LESB-II kaon beamline of the AGS at Brookhaven National Laboratory.
The beamline parameters are given in Table~\ref{tbl-LESBII}.
The experimental layout is shown in Fig.~\ref{layout}.
A beam of 750 MeV/c kaons incident on a liquid helium target produced
$^4_\Lambda{\rm He}$ through the $^4\text{He}(K^-,\pi^-)^4_\Lambda\text{He}$ 
reaction, and pions
from the production reaction were detected near $0^\circ$ 
in the  Moby Dick spectrometer.
Detector packages located on each side of the
target area were used to detect and identify pions, protons, neutrons, and
$\gamma$'s from the hypernuclear decays.

\begin{table}
\caption{\label{tbl-LESBII}LESB-II and Moby Dick Spectrometer.}
\begin{ruledtabular}
\begin{tabular}{l c}
Item & Description \\
\hline
\multicolumn{2}{c}{LESB-II Separated Beamline/Spectrometer} \\
Momentum     & 750~MeV/c\\
Momentum acceptance & 5\%~FWHM \\
$\pi :K$ ratio at target & $13:1$ \\
$K^-$ flux        &  $2\times 10^5 \text{s}^{-1}$ \\
Momentum measurement & Drift chambers ID1X -- ID7U \\
                     & and hodoscope MH\\
$\pi - K$ Particle ID & IS1 -- IT1,2 TOF scintillators\\
                      & and CP \Cerenkov $\pi$-veto \\
 & \\
\multicolumn{2}{c}{Moby Dick Spectrometer}\\
Momentum      & 680~MeV/c \\
Momentum acceptance  & 5\%~FWHM \\
Momentum measurement & Drift chambers ID9Y -- ID15Y \\
$\pi - K$  particle ID & IT1,2 -- IS2,3 TOF scintillators \\
\end{tabular}
\end{ruledtabular}
\end{table}

The incident kaon trajectories were determined with drift chambers located
just upstream of the target (ID2X through ID7U).
This information was combined with data from
the drift chamber ID1X, located upstream of the \Km spectrometer, to 
determine the incident kaon momentum.  Although the
kaon-to-pion ratio at the target was only 1:13, the kaons were cleanly
separated using a combination of the time-of-flight between scintillator IS1
(not shown) located upstream of the \Km spectrometer and scintillators
IT1 and IT2 and a critical angle \Cerenkov detector (CP).
The outgoing pion momentum was measured with
the Moby Dick spectrometer located at zero degrees downstream of the
target, and the outgoing particle identification was
accomplished with time-of-flight measurements.

The 30.5 cm long liquid helium target vessel consisted of a 6.35~cm
diameter Dacron cylinder with a wall thickness of 0.020 cm.  
To minimize material between the decay vertex and the out-of-beam decay
particle detectors,
the vacuum vessel
containing the Dacron shell was constructed from a PVC foam cylinder
of density 0.053~gm/cm$^3$ with inner and outer radii of 7.6~cm and
11.43~cm, respectively.

The separation of the decay protons and 
pions was accomplished
by combining the crude range information from all out-of-beam detectors
with the measured rate of energy loss
($dE/dx$), total energy loss, and the measured time-of-flight (TOF) 
between the Decay-Timing and TOF scintillator layers
(also referred to in this paper as OT1 and OT2, respectively).
The precision timing scintillators of the Decay-Timing layer also played an 
important part
in the determination of the lifetime of the \LfourHe hypernucleus.
With no magnetic field in the region of the out-of-beam detectors, the
detector system of E788 was unable to distinguish a \pim from a $\pi^+$,
making it possible to measure only the sum of the two widths \Gpim and 
$\Gamma_{\pi^+}$.
However, the width $\Gamma_{\pi^+}$ has been seen to be small compared to
the \pim decay width 
($\Gamma_{\pi^+}/\Gamma_{\pi^-} = 0.043 \pm 0.017$~\cite{Keyes76})
and was ignored in this analysis.
Large volume
($\approx1.4~{\rm m}^3$) neutron detector arrays were placed near the target
to directly measure the decay neutrons and were used to determine the 
neutron energies by time-of-flight.
The neutron detector arrays were also used to detect the $\gamma$'s from
the decay of $\pi^\circ$'s originating from \LfourHe decay.
Thus, this experiment directly measured $\Gamma_{\rm total}$, 
$\Gamma_{\pi^-}$, $\Gamma_{\pi^\circ}$, $\Gamma_p$, and $\Gamma_n$,
along with the kinetic energy spectra for the decay $\pi^-$'s, protons, 
and neutrons.
Details of the detector dimensions are given in Table~\ref{tbl-detectors}.

\begin{table*}
\caption{\label{tbl-detectors} Descriptions of the target area
detectors utilized for the hypernuclear decay measurement.
An overview of the layout of these detector elements is shown in 
Fig.~\ref{layout}. }
\begin{ruledtabular}
\begin{tabular}{l c c}
Name & Function & Dimensions (cm$^3$) \\
\hline
\multicolumn{3}{c}{Drift Chambers} \\
ID    & $K - \pi$ trajectories & $12.2 \times 12.2 \times 0.43$ \\
OD    & y and z track positions (1 pair each side) 
 & $30.5 \times 30.5 \times 1.9$ \\
 & & \\
\multicolumn{3}{c}{Scintillation Counters} \\
IT1, IT2  &$ K-\pi$ TOF and \LfourHe formation time 
 & $15.0 \times 4.0 \times 1.27$ \\
CP & critical angle pion \Cerenkov counter & $15.0 \times 4.0 \times 0.6$ \\
OT1   &decay time and $dE/dx$ (5 each side)  
 & $22.0 \times 4.5 \times 0.64$   \\
OT2   &TOF, $dE/dx$, and range (10 each side) 
 & $150.0 \times 12.0 \times 1.5$ \\
NV    &range and charged particle veto (9 each side) 
 & $182.9 \times 22.9 \times 0.95$ \\
NA    &charged particle veto (2 each side) 
 & $182.9 \times 22.9 \times 0.95$ \\
ND    &range and neutral particle detection (50 each side) 
 & $182.9 \times 15.2 \times 5.1$ \\
\end{tabular}
\end{ruledtabular}
\end{table*}

\section{Out-of-Beam Track Reconstruction and Identification}
\label{sec:pid}

The task of the out-of-beam tracking was to sort through the scintillator
hits recorded in the out-of-beam detector arrays and determine the 
trajectories of the charged and neutral particles that created them.
A hit in the out-of-beam detector array was defined
as a coincidence of signals in the photomultiplier tubes at each end of 
a scintillator element,
requiring ADC and TDC information for both of the PMTs.
Each hit was characterized by a hit time, energy deposition, 
and hit position.
For a hit to be considered for tracking, the recorded time (measured relative
to the event start-time as set by the mean time of 
in-beam scintillators IT1 and IT2) and 
measured energy deposition must have been greater than zero.

As a preliminary step, neutron detector hits that satisfied the rudimentary
conditions described above were examined, and correlated hits were grouped
into {\em clusters} based on relative hit times and positions
({\em i.e.}, groups of hits that were sufficiently close in time and space 
were considered to have arisen from a single particle).
The hit clusters were classified as {\em charged} if the group included
any hits in the OT2 or NV layers, otherwise the cluster was considered
to have originated from a neutral particle.


Charged tracks were indicated by a coincidence of hits in the OT1 
and OT2 layers.
The charged tracking began with a search for hits in the OT1 layer,
and for each OT1 hit found, a search was made for corresponding hits in OT2.  
If the hit pairs were properly time ordered ({\em i.e.}, the OT1 hit preceded the 
OT2 hit), 
they were considered to be part of a charged track and were then subjected
to a more detailed tracking procedure.
The final track was found from a linear least squares fit to 
the OT1 and OT2 hits, the \Kpi reaction vertex, out-of-beam drift chamber hits,
and any hits in the neutron veto layer or neutron detector array that 
belonged to the same cluster as the OT2 hit.
Interesting quantities such as the
particle's velocity between the OT1 and OT2 layers and
the particle's range and total energy loss in the out-of-beam
detector system were then determined.


The tracking information found above was used to distinguish 
protons from pions,
a task that was
complicated by the absence of a momentum-analyzing magnetic field.
To determine the identity of the charged particles,
the measured velocity, range, and energy loss of a charged track were
compared with the expected behavior of the hypernuclear decay products
via the quantity:
\begin{multline}
\label{eq:anal:dpid}
D_{\pi,p} = \bigg[ \left( \frac{\Delta E_{meas} 
  - \Delta E_{\pi,p}(\beta_{fit})}{\sigma_E}\right)^2 \\
  + \left( \frac{\beta_{meas} - \beta_{fit}}{\sigma_\beta} \right)^2 
    \bigg]^{\frac{1}{2}},
\end{multline}
where $\Delta E_{meas}$ is the measured energy deposition in the OT2
scintillator layer, $\beta_{meas}$ is the velocity fraction as 
measured in the time-of-flight region of the out-of-beam detector array,
and $\sigma_E \sim 1/\sqrt{E}$ and $\sigma_\beta \sim \beta^2$ 
are the resolutions for the measured quantities.
To find the quantity $\beta_{fit}$ (and thus, the theoretical energy 
deposition $\Delta E_{\pi,p}$), the above expression for $D_{\pi,p}$ was
minimized by varying the value of $\beta_{fit}$ within limits determined by
the measured range of the particle.
The PID (particle identification) for a given track was then determined by
testing the values of $D_\pi$ (calculated assuming the track was created by
a pion) and $D_p$ (calculated assuming the track was created by a proton)
against predetermined limits.
The particle was then classified as either a proton, a pion, or an 
{\it unknown} (neither $D_\pi$ nor $D_p$ was within the appropriate limits).
The unknown particles consisted mostly of electrons and were
well separated from the pions and protons.


A neutral particle appeared as a 
single hit or small cluster of hits in the out-of-beam detector arrays, 
providing little information for tracking.
A neutral {\em track} was taken as a straight line connecting the
\Kpi reaction vertex with the out-of-beam neutral cluster,
where the time and position of the reaction vertex were known from the in-beam 
tracking.
For each of the neutral clusters found in the preliminary grouping,
a cluster hit time, position, and energy were calculated.
The hit time and position were determined by the member hit that was closest
to the target, and the energy was taken as the sum
of the energy deposition for all member hits.
Additionally, each cluster was characterized by a cluster size defined 
simply as the total number of member hits.
Once the neutral track was determined by the production vertex and cluster
hit position, 
the velocity fraction of the neutral particle could be 
calculated.  

The next task was to separate neutrons from $\gamma$'s and attempt to 
suppress the large accidental neutral background.
To reduce accidentals, an energy deposition threshold of $E_{dep} > 5$~MeVee
(MeV {\em electron equivalent})
was applied to the neutral clusters, and clusters with corresponding hits in
the auxiliary veto counters NA were discarded.
Good neutron candidate events were then selected by requiring the neutral
cluster to have fewer than 3 member hits and a value for the neutron 
kinetic energy (as derived from the measured time-of-flight) 
from 50 to 200~MeV.
The lower bound at 50~MeV avoids the energy region where accidentals 
dominate, and the upper bound at 200~MeV, which is well above the maximum 
expected kinetic energy for neutrons from hypernuclear decay ($\sim$170~MeV), 
removes $\gamma$'s from consideration.
Good $\gamma$ events were selected by requiring the measured $\beta$ 
of the
candidate track to be in the range $0.666 < \beta < 2.0$, 
where the lower bound at 0.666 is well above the highest
$\beta$ expected for neutrons from hypernuclear decay.
Even with the cuts described above, a sizeable background still remained in
the neutron and $\gamma$ samples.
The subtraction of this background will be discussed in subsequent sections.
A cancellation effect inherent in the subtraction of the background
reduces the sensitivity of the final results to the selection of the cut 
parameters given above.

\section{$^{4}_\Lambda{\rm He}$ EXCITATION SPECTRA}
\label{sec:ee}

Reconstruction of the $K^-$ and $\pi^-$ trajectories through their respective
spectrometers allowed the calculation of the invariant mass of the unobserved
strange system.  
Unlike p-shell hypernuclei, the ground state is also a
substitutional state and thus the $(K^-,\pi^-)$ reaction near zero degrees
is ideal for creating $^{4}_\Lambda{\rm He}$. 
The main source of background for the production reaction was in-flight
kaon-decays.
Those decays that may satisfy the $(K^-,\pi^-)$ trigger condition
include  
(i) $K^- \rightarrow \pi^- \pi^\circ$, 
(ii) $K^- \rightarrow \mu^- \bar{\nu}_\mu$, and
(iii) $K^- \rightarrow \mu^- \bar{\nu}_\mu \pi^\circ$.
For kaons that decayed within the area of the target,
the two-body decays were easily removed using the decay kinematics,
while the three-body decays had to be modeled with a Monte Carlo simulation
and fit to the data.
Some kaons decayed within the beamline chambers or Moby Dick
spectrometer or traversed the experimental area completely.
These events were characterized by a small apparent \Kpi scattering angle 
and were removed by requiring an angle greater than 25~mrad.
Events with production vertices outside the target region or poor beam track
reconstruction were also rejected.

\begin{figure}[ht]
\includegraphics[width=8.6 cm,clip]{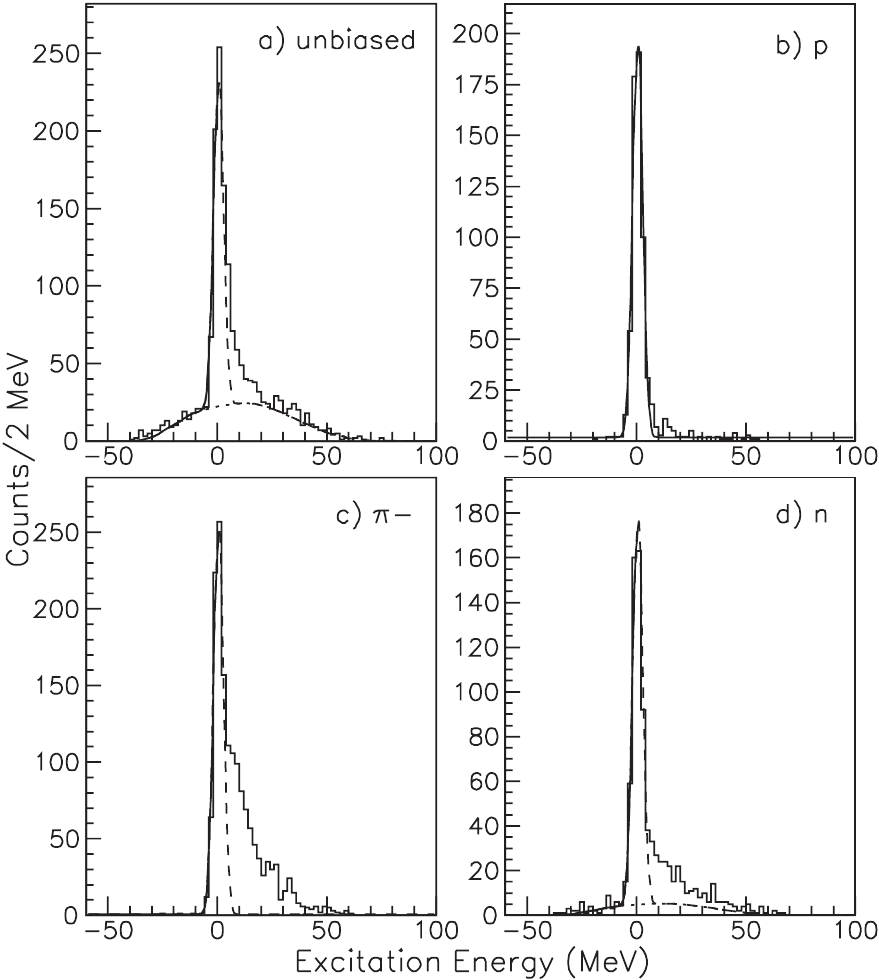}
\caption{
The excitation energy spectrum for $^4\rm{He}(K^-,\pi^-)X$ is shown for
several cases:
(a) unbiased by out-of-beam tags, (b) tagged by a proton in the
out-of-beam detector package, (c) tagged by a pion, and (d) tagged by
a neutron.  
Overlaid on each plot is the result of the fit to the ground-state peak
(including appropriate backgrounds as described in the text).
\label{ee}}
\end{figure}

The resulting excitation energy spectrum for unbiased $(K^-,\pi^-)$ 
events is shown in Fig.~\ref{ee}(a), 
where {\em unbiased} refers to the trigger condition in which only an
incoming \Km and outgoing \pim were required ({\em i.e.}, no information from the 
out-of-beam detectors was considered).
Also shown are the excitation energy spectra for events with a coincident
proton, pion, or neutron [Figs.~\ref{ee}(b), (c), and (d), respectively]
where the coincident decay particles were identified as described in the
previous section.
(Note that the low number of counts in the unbiased spectrum as compared
to the coincident spectra is a result 
of a hardware prescale in the unbiased trigger.)
Since the coincident protons were required to have
kinetic energies greater than 35~MeV, only events in which the $\Lambda$
was bound to a nucleus and decayed through a nonmesonic channel contribute
to the proton-tagged spectrum of Fig.~\ref{ee}(b).  
A fit of a Gaussian shape to this peak determined an
experimental resolution of $2.3$~MeV RMS and defined the zero of excitation
energy.  The unbiased spectrum, Fig.~\ref{ee}(a), was then determined to
consist of three parts: (i) a broad background from \Kmnp contamination whose 
shape was determined by Monte Carlo simulation
and whose magnitude was found from a fit to the lower part of the spectrum, 
(ii) events
corresponding to the production of the bound $^{4}_\Lambda{\rm He}$ ground
state with mean and width determined from Fig.~\ref{ee}(b) and amplitude
determined by a fit to the left
half of the peak, and (iii) additional structure above zero excitation energy
that we associate with the production of unbound $\Lambda$ hyperons.

The excitation energy spectrum for coincident decay pions of Fig.~\ref{ee}(c) is
shown with a two parameter fit to the left half of the ground-state peak.
The fit determines the amplitude of the ground-state peak and includes
a flat background term.
The requirement of a decay-pion tag virtually eliminates the residual
\Kmnp contamination but retains events from both
bound and unbound $\Lambda$ production as would be expected.
For the neutron-tagged excitation energy spectrum of Fig.~\ref{ee}(d), 
the coincident neutrons consist of a mixture of neutrons
from nonmesonic \LfourHe decay, accidental neutrons, and neutrons 
created when $\pi^-$'s from \LfourHe decay or quasifree $\Lambda$ decay 
were absorbed in the material around the target area.

\section{Lifetime of  $^{4}_\Lambda{\rm He}$}
\label{sec:life}

The lifetime of a \LfourHe hypernucleus was defined event-by-event as the 
time between
the formation of the hypernucleus and its subsequent decay as determined
from the velocity and trajectory of the incident kaon and
the energy and trajectory of the charged hypernuclear decay product. 
The \LfourHe hypernucleus was assumed to be created at rest, and 
the position of the reaction vertex (and decay point) was 
inferred from the in-beam and
out-of-beam tracks using a distance of closest approach (DCA) algorithm.
The velocity of the incoming \Km was derived from its momentum as measured by
the upstream \Km spectrometer, and its trajectory was determined
by the six drift chambers located just upstream of the \fourHe target.
A time measurement was also made for the beam kaon as it passed the in-beam
scintillators IT1 and IT2.
This measurement set the start time for the event and was taken as
the average of the two counters (producing a combined resolution of 
$\bar{\sigma}_{\text{IT}} = 44.3$~ps).
The time of hypernuclear formation was then simply
found from the velocity of the $K^-$ and its path length from the 
point midway between the
IT1 and IT2 timing counters to the reaction point.
(The velocity of the beam kaon was assumed to have remained constant as it 
traversed the experimental area.)
For the out-of-beam charged track, the energy was determined from the
time-of-flight as measured between the Decay-Timing (OT1) and TOF (OT2) 
scintillator layers.
The decay time at the vertex position was determined after making corrections
for the unmeasured energy losses in the target region.

\begin{figure}[b]
\includegraphics[width=8.6 cm,clip]{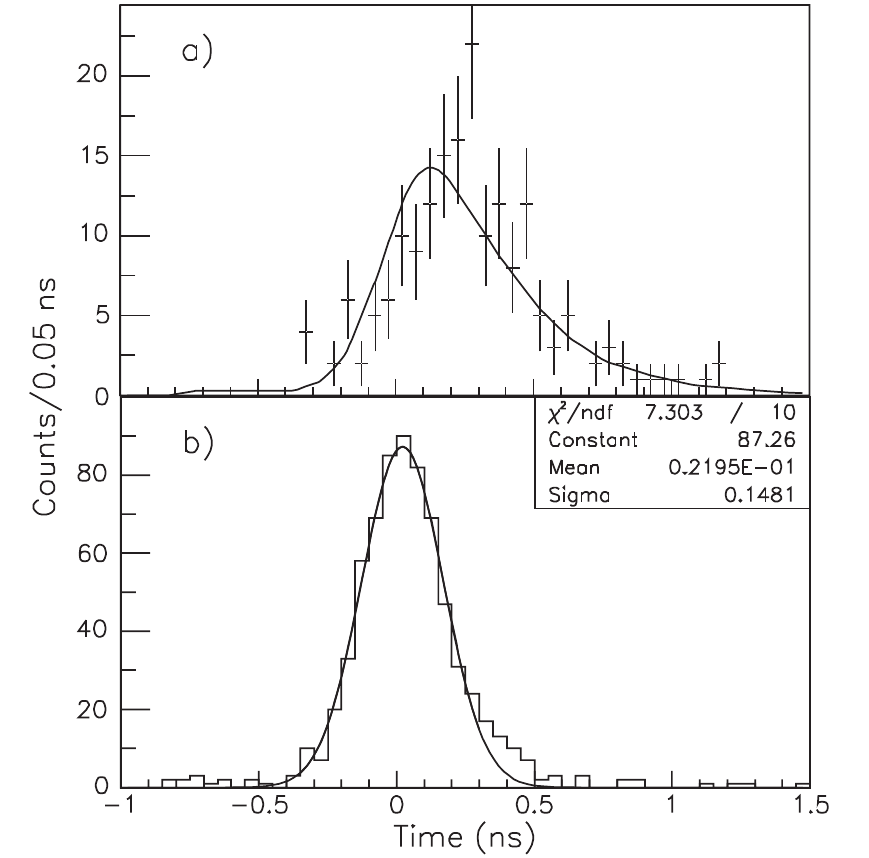}
\caption[]{\label{fig:life}
Figure (a) shows the \LfourHe lifetime distribution for events with a 
coincident
proton overlaid with the fit of the expected distribution.
Figure (b) shows the time-difference distribution for protons from the reaction
\fourHe$(\pi^-,p)X$ used to determine the time resolution function.
}
\end{figure}

The events used in the \LfourHe mean lifetime determination were subject
to the in-beam tracking cuts described in Sec.~\ref{sec:ee} for the 
hypernuclear excitation energy spectra.
Protons were chosen in the out-of-beam to avoid the contamination of 
quasifree $\Lambda$ decay events present in the $\pi^-$-tagged sample.
Also,
protons with kinetic energy below 70~MeV were not considered for the lifetime
measurement
since the uncertainty in the energy loss correction increases at lower
energies.
To ensure that the kaon and proton tracks were correlated with a
hypernuclear formation event:
(i) the excitation energy was required to be within $\pm 2\sigma$ of the mean,
(ii) the \Km and proton tracks must have had a DCA of less than 2~cm, and
(iii) the reaction vertex must have been within the liquid \fourHe containment 
vessel.
The resulting lifetime distribution for these proton-tagged events can be 
seen in Fig.~\ref{fig:life}(a).

The mean lifetime of \LfourHe was determined by a fit to the distribution of
Fig.~\ref{fig:life}(a).
The actual time distribution expected for this measurement has the 
form:
\begin{equation}
\label{eq:life}
L_\tau(t) = \int^{\infty}_{-\infty} dt' R(t') P_\tau(t - t'),
\end{equation}
where $P_\tau(t)$ is the decay probability distribution:
\begin{equation}
P_\tau(t) = \frac{1}{\tau} e^{-t/\tau} \theta(t),
\end{equation}
and $R(t)$ is the resolution function as determined from the time 
distribution of prompt protons originating from the reaction 
$\pi^- + ^4\text{He} \rightarrow p + X$ [shown in Fig.~\ref{fig:life}(b)]
and $\theta(t)$ is the usual theta function.
The measured distribution of Fig.~\ref{fig:life}(b) gives a resolution of 
$\sigma_\tau \simeq 150$~ps.
The fit to the hypernuclear lifetime distribution of Fig.~\ref{fig:life}(a) was 
performed using a likelihood function derived assuming Poisson statistics 
from the function:
\begin{equation}
\label{eq:lfit}
f(t) = C_1^2 + C_2 t + C_3^2 L_\tau(t),
\end{equation}
which includes two background terms in addition to the expected lifetime
distribution of Eq.~\ref{eq:life}.
By maximizing the likelihood function on the four parameter space 
($C_1$, $C_2$, $C_3$, $\tau$),
the value for the mean lifetime of \LfourHe was found to be 
$\tau = 245 \pm 24$~ps.
[The values for the background terms $C_1$ and $C_2$ were found
to be negligible compared to $C_3^2 L_\tau(t)$.]

\section{Particle Emission Spectra}
\label{sec:ke}

\begin{figure}[ht]
\includegraphics[width=8.6 cm,clip]{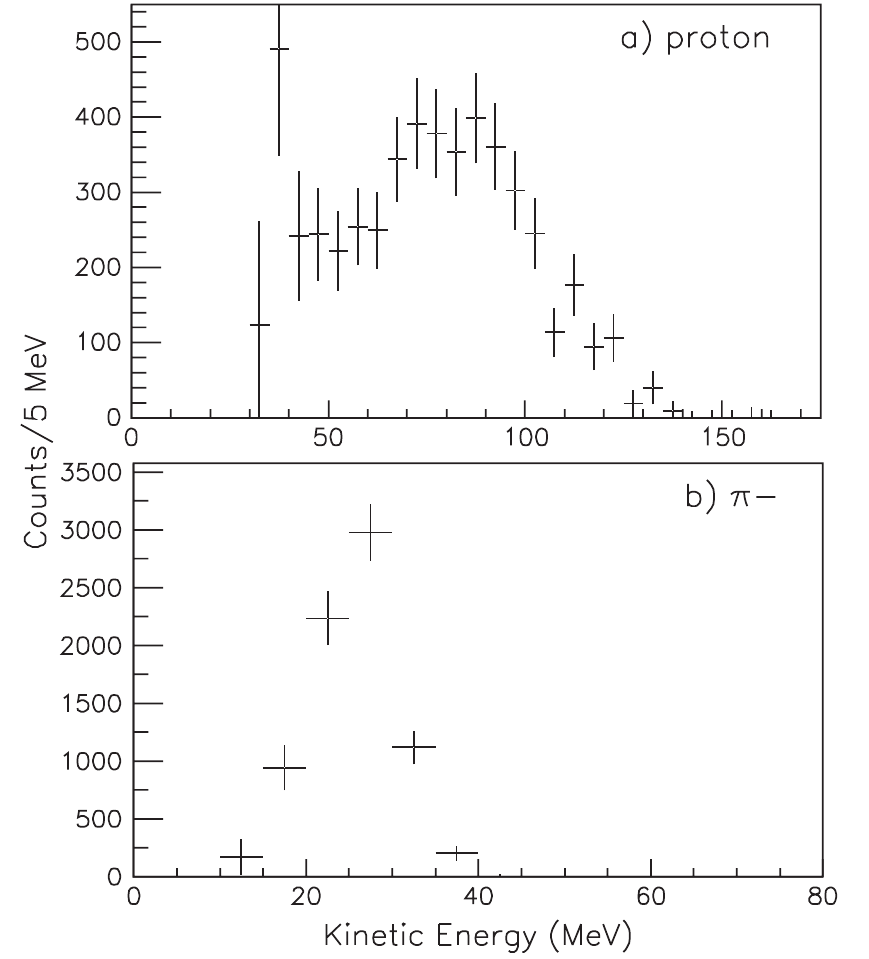}
\caption[]{\label{fig:keq}
Shown here are the
kinetic energy spectra for (a) protons and (b) pions detected in coincidence
with \LfourHe formation.
These spectra have been corrected for PID inefficiencies, detector acceptance,
and unmeasured energy losses in the target region.}
\end{figure}

\begin{figure}[t]
\includegraphics[width=8.6 cm,clip]{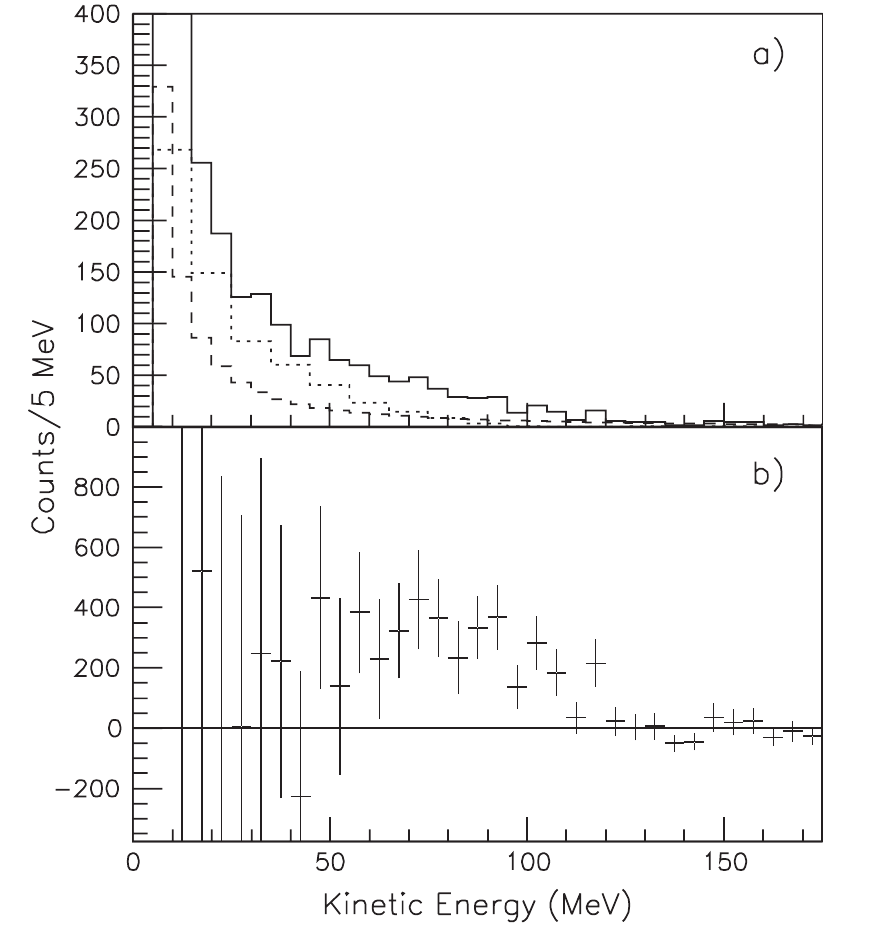}
\caption[]{\label{fig:ken}
Figure (a) shows the raw neutron kinetic energy spectrum with background 
contributions from 
accidental neutrons (dashed line) and neutrons from \pim absorption
(dotted line) overlaid.
Figure (b) shows the neutron spectrum after background subtraction and
corrections for detector acceptance.}
\end{figure}

The kinetic energy spectra for the decay protons, pions, and neutrons are 
shown in 
Figures~\ref{fig:keq}(a), \ref{fig:keq}(b), and \ref{fig:ken}(b), respectively.
These events were subject to the cuts described in Sec.~\ref{sec:ee} for the
\LfourHe excitation energy spectra.
To reduce contamination from out-of-beam particles not originating from 
a hypernuclear ground-state event, 
a cut was also placed on the value of the excitation energy.
In the case of proton- and neutron-tagged events, the excitation energy, 
$E \negthinspace E$, was
required to be within $\pm 2\sigma$ of the mean of the ground-state peak,
while in the case of pion-tagged events, a tighter cut of 
$\mu-2\sigma < E \negthinspace E < \mu$ (referred to here as the 
``$-2\sigma$ cut'')
was employed to reduce contamination from quasifree $\Lambda$ decay events.
(The quantities $\mu$ and $\sigma$ are the previously determined mean and width
of the \LfourHe ground-state peak.)
For the proton and pion spectra, corrections for the unmeasured
energy losses in the target region have been included.

In order to obtain
the neutron kinetic energy spectrum of Fig.~\ref{fig:ken}(b), a 
considerable background
consisting of accidental neutrons and neutrons from \pim interactions had
to be subtracted from 
the raw neutron kinetic energy spectrum shown in Fig.~\ref{fig:ken}(a).
The expected shape of the accidental neutron spectrum was generated assuming 
a flat distribution in $\beta^{-1}$ ({\em i.e.}, a flat time distribution).
The distribution for neutrons from \pim absorption was then extracted from
the data by considering neutron-tagged events with excitation energy 
in the range $10 < EE < 30$~MeV.
The neutrons from these events consist of a mixture of
accidental neutrons and neutrons from \pim absorption only.
The normalizations for the background shapes were found via a detailed 
comparison of the unbiased and neutron-tagged excitation spectra of 
Figures~\ref{ee}(a) and \ref{ee}(d).
The background contributions thus determined are overlaid on the plot of 
Fig.~\ref{fig:ken}(a).

The kinetic energy spectra of Figures~\ref{fig:keq}(a), \ref{fig:keq}(b), and
\ref{fig:ken}(b) also include corrections for small PID inefficiencies 
and detector acceptance which were derived with the aid of a Monte Carlo
simulation of the E788 out-of-beam detector system.
Protons, neutrons, and $\pi^-$'s with realistic initial energy distributions
were propagated through the target area and out-of-beam detector system.
The response of the scintillator counters was modeled using an approximate
form of the Bethe-Bloch equation of energy loss in the case of charged 
particles
and the DEMONS software package~\cite{Byrd92} for neutron interactions. 
Unmeasured energy losses in the target materials and detector wrappings,
pion decays and interactions, and
corrections for the non-linear response of the scintillator
material at large energy deposition were also included.
The tracking and particle identification was performed with the same code as
used for the real data, and the results were used to derive the necessary
corrections.

\section{Extraction of Partial Rates}
\label{sec:rates}

The decay modes considered in this analysis were the
$\pi^\circ$, $\pi^-$, proton-stimulated, neutron-stimulated, 
and three-body nucleon-stimulated modes with branching fractions defined as:
\begin{equation}
\label{eq:bratio:bf}
B_a = \frac{\Gamma_a}{\Gamma_{tot}} = \frac{N_a}{N^{tot}_{\text{HN}}},
\end{equation}
where $a=\pi^\circ,\pi^-,p,n$, and $mb$;
$N_a$ is the total number of hypernuclear events decaying to mode $a$; 
and $N_{\text{HN}}^{tot} = 27800 \pm 1500$ is the total number of \LfourHe 
hypernuclear events found from the fit to the ground-state peak of the 
unbiased excitation spectrum of Fig.~\ref{ee}(a).
This total also includes corrections for 
a hardware prescale in the unbiased trigger.
The \piz branching fraction was extracted by considering $\gamma$-tagged
hypernuclear formation events, while
the \pim branching fraction was determined from 
the observed \pim kinetic energy distribution.
The nonmesonic decay branching fractions were extracted by two essentially
independent methods utilizing
(i) the single-particle kinetic energy spectra with final-state interactions
and three-body \LNN decay contributions considered within a simple model, and
(ii) multiple nucleon coincidence events.
The final values for the \LfourHe branching fractions were then determined
by a $\chi^2$ minimization that considers the mesonic and nonmesonic rates
simultaneously.

\subsection{The \piz Decay Mode}
\label{sec:piz}

The \piz decay rate was found by considering events for which one or more 
$\gamma$'s were detected in coincidence with \LfourHe hypernuclear formation.
The $\gamma$'s were identified as described in Sec.~\ref{sec:pid}, and
the candidate events were subject to the same cuts as the excitation 
spectra of Sec.~\ref{sec:ee}.
Three possible sources of $\gamma$-rays were considered for the \piz rate
extraction: \piz decay, \pim interactions, and accidental hits.
Thus, the possible backgrounds include $\gamma$-tagged events arising from 
accidentals, 
$\pi^-$'s and $\pi^\circ$'s from quasifree $\Lambda$ decay, 
$\pi^-$'s from \LfourHe decay, 
and $\pi^\circ$'s from the \Kmnp decay background.
To find the \piz branching fraction, these background events were 
subtracted as described below.

In order to determine the quantitative contributions of each type of 
background,
the various sources of $\gamma$'s were isolated by considering $\gamma$-tagged
events with particular excitation energies.
First, 
events with excitation energy below $-10$~MeV, 
which consist almost exclusively of 
accidental $\gamma$'s and $\gamma$'s from the $K^-$ decay background,
were used to estimate the detection efficiencies for accidentals
and $\gamma$'s from \piz decay.
Next, the $\gamma$'s from \pim interactions were studied
using events with excitation energy in the range 
$10 < E \negthinspace E < 30$~MeV.
These events consist of accidental $\gamma$'s and $\gamma$'s from the 
$K^-$ decay background (as in the previous case) with additional 
background $\gamma$ events originating from quasifree $\Lambda$ decay.
Using the accidental and \piz decay $\gamma$ detection probabilities
determined above,
the contribution from 
quasifree $\Lambda \rightarrow p \pi^-$ decay could be extracted,
allowing the determination of the detection probability for $\gamma$'s from
\pim interactions.
Finally, the \LfourHe \piz decay events were selected by requiring the 
excitation
energy to be within $\pm 2\sigma$ of the ground-state peak and subtracting the
background contributions found above.

After properly accounting for the backgrounds and 
$\gamma$ detection efficiency, 
the \piz branching fraction was found to be
$B_{\pi^\circ} = 0.552 \pm 0.076 \text{(stat.)} \pm 0.061 \text{(syst.)}$.
So, for a $\Lambda$ embedded within the $_\Lambda^4$He hypernucleus, 
the \piz decay rate:
\begin{equation}
\frac{\Gamma_{\pi^\circ}}{\Gamma_\Lambda} 
= B_{\pi^\circ} \times \frac{\Gamma_{tot}}{\Gamma_\Lambda}
= 0.59 \pm 0.10,
\end{equation}
expressed here in units of the total decay rate of the free $\Lambda$,
is enhanced by 
a factor of about 1.6 as compared to the free $\Lambda$ decay partial width
of $\Gamma_{\pi^\circ}^{free}/\Gamma_\Lambda = 0.358 \pm 0.005$~\cite{PDG06}.
(The quantity $\Gamma_{tot}/\Gamma_\Lambda = 1.07 \pm 0.11$ was derived from
the \LfourHe hypernuclear lifetime found in Sec.~\ref{sec:life}).

\subsection{The \pim Decay Mode}
\label{sec:pim}

\begin{figure}[ht]
\includegraphics[width=8.6 cm,clip]{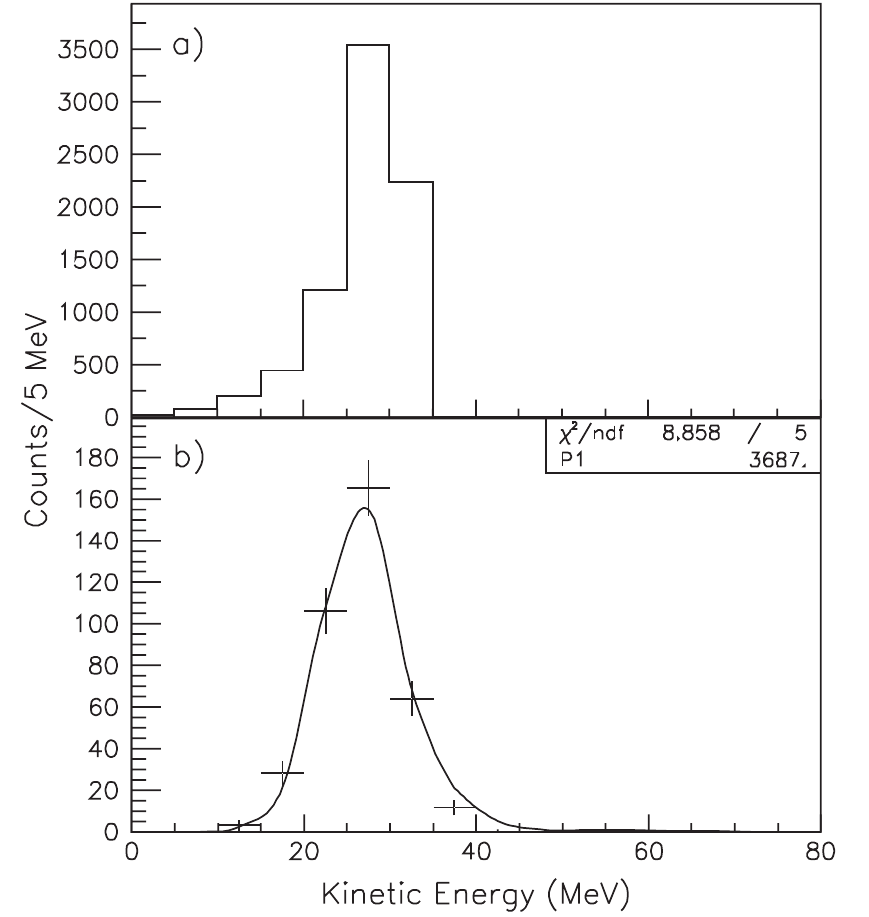}
\caption[]{\label{fig:pifit}
Shown here are (a) the theoretical \pim kinetic energy spectrum of
Kumagai-Fuse, {\em et al.}~\cite{Izumi96} and (b) a fit of this expected shape
to the measured distribution after including the effects of detector 
acceptance and resolution.
The measured spectrum of figure (b) has been corrected for PID inefficiencies
and unmeasured energy losses (but not detector acceptance).
}\end{figure}

The \pim decay rate was determined by a fit to the observed \pim
kinetic energy spectrum using a theoretical distribution provided by 
the model calculation of Kumagai-Fuse, {\em et al.}~\cite{Izumi96}
Their predicted \pim spectrum [shown in Fig.~\ref{fig:pifit}(a)]
was derived using the resonating group method and includes
final-state interactions and pion distortion effects.
The effects of detector acceptance and resolution unique to E788 were
applied to the model by using the spectrum of Fig.~\ref{fig:pifit}(a) as 
input to the Monte Carlo simulation described in Sec.~\ref{sec:ke}
and reconstructing the initial kinetic energy distribution with
the analysis code.

A fit of the model prediction to the PID-corrected \pim spectrum 
of Fig.~\ref{fig:pifit}(b) was then
performed using the fit function:
\begin{equation}
\label{eq:acc:pifit}
N_\pi(E) = M_\pi^{2\sigma} h_\pi(E),
\end{equation}
where $h_\pi(E)$ is a parameterization of the expected \pim kinetic energy 
distribution
as reconstructed by the analysis code (normalized to the detector acceptance
and in units of counts/bin per event) and
$M_\pi^{2\sigma}$ is the fit parameter representing the acceptance-corrected 
number of pions within the $-2\sigma$ excitation energy cut.
The result of the fit is shown in Fig.~\ref{fig:pifit}(b) where
the value of $M_\pi^{2\sigma}$ was found to be $3690 \pm 190$.
Correcting for the excitation energy cut yields a value for the total number
of $\pi^-$'s:
\begin{equation}
M_\pi^{total} = g_\pi M_\pi^{2\sigma} = 7490 \pm 480,
\end{equation}
where $g_\pi = 2.030 \pm 0.076$ is the ratio of the number of counts in the
ground-state peak of the $\pi^-$-tagged excitation spectrum of 
Fig.~\ref{ee}(c) to the number of events within the $-2\sigma$ cut.

The $\pi^-$-decay branching fraction for the $_\Lambda^4$He hypernucleus 
was then found by Eq.~\ref{eq:bratio:bf} as 
$B_{\pi^-} = 0.269 \pm 0.022 \text{(stat.)} \pm 0.014 \text{(syst.)}$.
In the present case of $_\Lambda^4$He decay, the observed \pim decay rate:
\begin{equation}
\frac{\Gamma_{\pi^-}}{\Gamma_\Lambda} 
 = B_{\pi^-} \times \frac{\Gamma_{tot}}{\Gamma_\Lambda}
 = 0.289 \pm 0.037,
\end{equation}
in units of the free $\Lambda$ decay,
is seen to be suppressed by a factor of about 2.2 as compared to the free 
$\Lambda$ decay value of 
$\Gamma_{\pi^-}^{free}/\Gamma_\Lambda = 0.639 \pm 0.005$~\cite{PDG06}.
This suppression is mainly due to the Pauli blocking of the final-state proton.

\subsection{Nonmesonic rates from single-particle KE spectra}
\label{sec:nmes}

\begin{figure}[t]
\includegraphics[width=8.6 cm,clip]{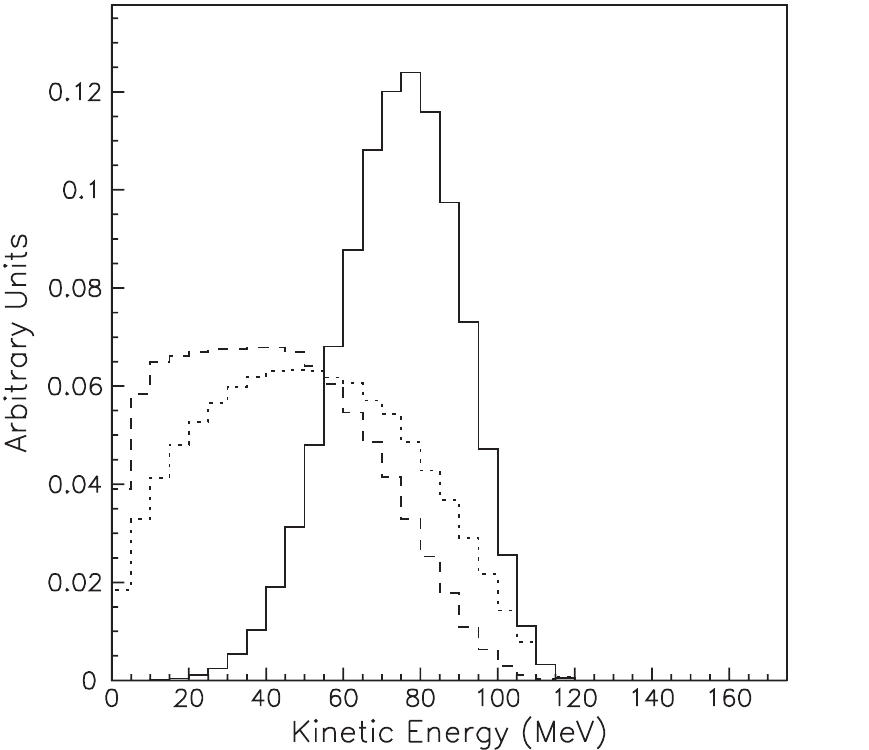}
  \caption[]{ 
Shown here are the predicted no-scatter and rescatter kinetic 
energy distributions for nucleons 
from \LN decay (solid and dashed lines, respectively) along with the
distribution for the primary nucleons from \LNN decay (dotted line).
\label{fig:pmod} }
\end{figure}

The rates for proton- and neutron-stimulated decay were extracted
simultaneously from the observed proton and neutron kinetic energy spectra
using a simple model that included the effects of final-state interactions
(FSI).
In the model,
two types of decay events were considered: 
\LN decay with no FSI (referred to here as {\em no-scatter}) and 
decay events where one of the decay nucleons scatters from one of the
spectator nucleons ({\em rescatter}).
The expected no-scatter and rescatter nucleon kinetic energy distributions
(shown in Fig.~\ref{fig:pmod}) were derived from a Monte Carlo simulation 
of the \LN interaction.
(Also shown is the distribution for nucleons from \LNN decay, which will be
discussed later in this section.)
For the Monte Carlo events, the momenta of the four initial-state baryons
were generated from a Gaussian distribution with an RMS width of 125~MeV/c
for the nucleons and 50~MeV/c for the $\Lambda$.
The distributions were correlated such that the sum of the four momenta
added to zero, and
the energy of each spectator nucleon was determined by setting its mass to 
be on-shell.
For no-scatter events,
the energy distribution for the primary decay nucleons 
was determined by choosing a direction
for the participating nucleon isotropically in the $\Lambda$-nucleon 
center-of-mass frame and boosting to the laboratory system.
The re-scatter events were 
simulated by
choosing a direction for one of the nucleons isotropically in the
decay-spectator nucleon system (s-wave) and boosting to the laboratory frame.

As in the \pim case above, 
the expected kinetic energy distributions were fit to the observed spectra, 
and the results were combined with the model to extract the
nucleon-stimulated decay rates.
To include the effects of the resolution and acceptance of the detector
system, the expected no-scatter and rescatter distributions of 
Fig.~\ref{fig:pmod} were used as input to the Monte Carlo simulation discussed
in Sec.~\ref{sec:ke} and then reconstructed using the analysis code.
The resulting spectra were designated $h^0(E)$ for
the nucleons that did not undergo FSI and $h^1(E)$ for the rescattered 
nucleons.
(These distributions differ for protons and neutrons due to differing energy
resolutions and detection acceptance.)
The distributions are in units of counts/bin per event and are normalized
to the detector acceptance.
The fits were then performed using functions of the form:
\begin{equation}
\begin{array}{lcccc}
N_a(E) &=& \underbrace{M_a^0 h_a^0(E)} &+& \underbrace{M_a^1 h_a^1(E)},\\
       & & \text{no-scatter} & & \text{rescatter}
\end{array}
\end{equation}
where $a = p, n$ and the fit parameters $M_a^0$ and $M_a^1$ represent the 
acceptance-corrected numbers of no-scatter and rescatter nucleons, 
respectively.
The fit parameters $M_a^0$ and $M_a^1$ were related via our model as
described below.

\begin{figure}[ht]
\includegraphics[width=8.6 cm,clip]{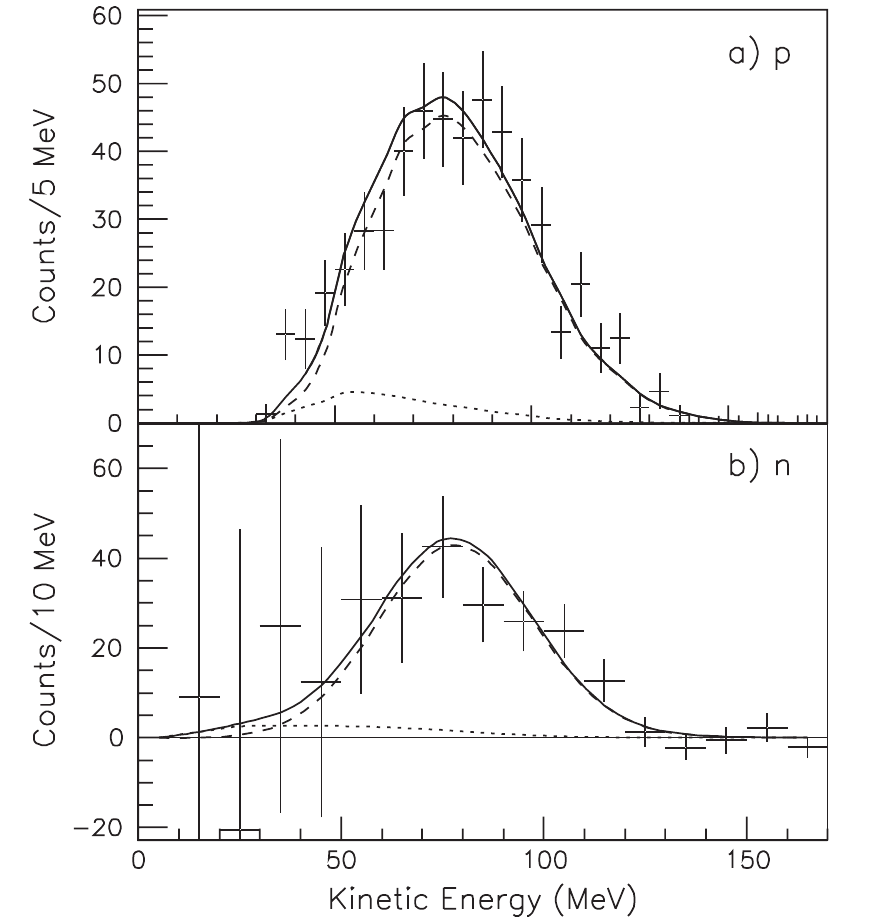}
  \caption[]{ 
Shown here are the fits of the expected no-scatter and rescatter distributions
to the observed (a) proton and (b) neutron kinetic energy spectra.
The no-scatter contribution is indicated by a dashed line, while the
rescatter contribution is shown as a dotted line.
\label{fig:fsi} }
\end{figure}

The effects of final-state interactions were incorporated by defining a 
rescatter probability, $\epsilon$, as the probability for one of the 
primary-decay nucleons from \LN decay to scatter off of one of the spectator
nucleons.
The value of $\epsilon$ was determined from the data as described below.
The scattering processes 
were assumed
to occur with equal probability
({\em i.e.}, $\epsilon$ has the same value for all proton-neutron rescatter 
combinations).
Since the effects of the FSI correction were seen to be small, 
this assumption was considered adequate.
Assigning a probability to each possible final state in terms of the
rescatter probability per spectator nucleon, $\epsilon$,
and the nucleon-stimulated branching fractions, $B_p$ and $B_n$, 
then summing over no-scatter and rescatter nucleons yields:
\begin{eqnarray}
M_p^0 &=& (1 - 2\epsilon) B_p N_{\text{HN}}^{tot} \label{eq:mp0}, \\
M_n^0 &=& (1 - 2\epsilon) B_p N_{\text{HN}}^{tot} 
       + 2 (1 - 2\epsilon) B_n N_{\text{HN}}^{tot} \label{eq:mn0}, \\
M_p^1 &=& 4 \epsilon (B_p + B_n) N_{\text{HN}}^{tot} \label{eq:mp1}, \\
M_n^1 &=& 4 \epsilon (B_p + B_n) N_{\text{HN}}^{tot} \label{eq:mn1}.
\end{eqnarray}
Using these expressions, the parameters $M_p^1$ and $M_n^1$ were eliminated
from the fit functions:
\begin{eqnarray}
N_p(E) &=& M_p^0 h_p^0(E) + \frac{2\epsilon}{1 - 2\epsilon} 
  (M_p^0 + M_n^0) h_p^1(E), \\
N_n(E) &=& M_n^0 h_n^0(E) + \frac{2\epsilon}{1 - 2\epsilon} 
  (M_p^0 + M_n^0) h_n^1(E).
\end{eqnarray}
The three parameters $M_p^0$, $M_n^0$, and $\epsilon$ were then determined
by a fit to the data (shown in Fig.~\ref{fig:fsi}).
Since the fit shapes $h^0(E)$ and $h^1(E)$ include the effects of 
the detector efficiency and resolution, the observed nucleon kinetic energy 
spectra used for the fits were not efficiency corrected.
The fits were carried out simultaneously by varying the quantity $\epsilon$
to obtain a minimum combined $\chi_2$.
Using the fit values, the nucleon-stimulated branching fractions were then
calculated using Eqs.~\ref{eq:mp0} and \ref{eq:mn0}, and the results
are listed in Table~\ref{tab:nmd}.
The upper limits for the neutron-stimulated branching fraction and the ratio
\Gnpr were determined by the prescription of Feldman and 
Cousins~\cite{Feldman98}, which defines the confidence levels in a consistent
way while imposing the physical boundary at zero. 
The upper limits are given at the 95\% CL.
Also listed are the results for the case of no FSI found by fixing the
rescatter probability $\epsilon$ to zero.

\begin{table*}
\caption{\label{tab:nmd} 
Results of the nonmesonic decay rate extraction considering FSI and \LNN 
decay contributions.
The decay rates are given in units of $\Gamma_{tot}$, and quoted errors are
statistical only.
The upper limits given for $\Gamma_n/\Gamma_{tot}$ and \Gnpr 
(listed directly beneath the associated quantities) 
are at the 95\% confidence level and include systematic errors. }
\begin{ruledtabular}\begin{tabular}{lccccc}
 & No FSI or \LNN & FSI & \LNN & FSI + \LNN & Systematic Error \\
\hline
$\chi^2$/d.o.f. & 0.78 & 0.72 & 0.69 & 0.71 & \\
$\epsilon$ & & $0.045 \pm 0.035$ & & $0.030^{+0.024}_{-0.021}$ & \\
$\Gamma_{mb}/\Gamma_{tot}$ & & & $0.042 \pm 0.028$ 
 & $0.017^{+0.015}_{-0.012}$ & \\
$\Gamma_p/\Gamma_{tot}$ & $0.158 \pm 0.011$ & $0.160 \pm 0.015$
 & $0.134 \pm 0.026$ & $0.150 \pm 0.016$ & $\pm 0.0089$ \\
$\Gamma_n/\Gamma_{tot}$ & $-0.0066 \pm 0.0060$ & $-0.0089 \pm 0.0075$ 
 & $-0.0133^{+0.0089}_{-0.0071}$ & $-0.0112^{+0.0084}_{-0.0077}$ 
 & $\pm 0.0104$ \\
 & $\leq 0.018$ & $\leq 0.017$ & $\leq 0.015$ & $\leq 0.017$ & \\
$\Gamma_{nm}/\Gamma_{tot}$ & $0.1510 \pm 0.0099$ & $0.151 \pm 0.013$
 & $0.162 \pm 0.020$ & $0.156 \pm 0.017$ & $\pm 0.0096$ \\
\Gnpr & $-0.042 \pm 0.037$ & $-0.056^{+0.046}_{-0.043}$
 & $-0.100 \pm 0.068$ & $-0.075 \pm 0.055$ & $\pm 0.063$ \\
 & $\leq 0.10$ & $\leq 0.10$ & $\leq 0.097$ & $\leq 0.098$ & \\
\end{tabular}
\end{ruledtabular}\end{table*}

The model was also extended to include possible contributions from three-body
\LNN decays.
The expected kinetic energy distribution for these nucleons
was generated using the same Monte Carlo code as for the \LN interaction
described above, and the resulting distribution is shown in 
Fig.~\ref{fig:pmod}.
In this case, the momentum of the single on-shell spectator was initially
selected, fixing the momentum and energy of the participating \LNN system.
Final-state momenta for these three baryons were then selected in the
system's center-of-mass frame according to a phase-space distribution 
generated by the CERNLIB GENBOD code.
The nucleon momenta were then boosted to the laboratory frame.

As in the case of FSI above, the contribution of \LNN decay, 
$\Gamma_{mb}/\Gamma_{tot}$, was determined by a fit to the observed
kinetic energy spectra of Fig.~\ref{fig:fsi}.
Because the \LNN decay nucleons and rescattered $\Lambda N$ nucleons are 
indistinguishable at the level of statistics of the current measurement, 
the fits were performed with an
$h^1(E)$ that consisted of a linear combination of the \LNN and rescatter
distributions.
The results for the nucleon-stimulated branching fractions are shown in
Table~\ref{tab:nmd} for the cases of (i) \LNN decay with no FSI for the \LN
decays ({\em i.e.}, $\epsilon = 0$) and 
(ii) a mixture of \LNN decay and FSI with equal strengths ({\em i.e.}, the quantity
$M_p^1$ consisted of a 50/50 mixture of \LNN decay events and rescattered 
\LN decay events).
From the case of FSI, the upper limit for the contribution
of final-state interactions to the total \LfourHe decay rate was found as
$\Gamma_{nm}^{\text{FSI}}/\Gamma_{tot}
 = 4 \epsilon (\Gamma_{nm}/\Gamma_{tot}) \leq 0.11$ (95\% CL).
Also, the multibaryon decay branching fraction found for the case of 
\LNN decay with no FSI gives an upper limit of 
$\Gamma_{mb}/\Gamma_{tot} \leq 0.097$ (95\% CL) 
for the contribution of the \LNN decay process.

\subsection{Nonmesonic rates from multiple nucleon coincidence data}

The \LN interaction produces two energetic nucleons that may
be detected in the out-of-beam detector arrays resulting in both 
proton--neutron ($pn$) and neutron--neutron ($nn$) coincidence events.
The observation of such events gives another 
window onto the nucleon-stimulated decay process.
In the present experiment, a total of 87 $pn$ coincidence events and
19 $nn$ coincidence events were observed.
These events were subject to the same in-beam 
tracking cuts as the excitation spectra of Sec.~\ref{sec:ee} along with
a $\pm 2\sigma$ excitation energy cut.
The protons and neutrons were identified in the out-of-beam detector arrays
as described in Sec.~\ref{sec:pid}.

\begin{figure}[ht]
\includegraphics[width=8.6 cm,clip]{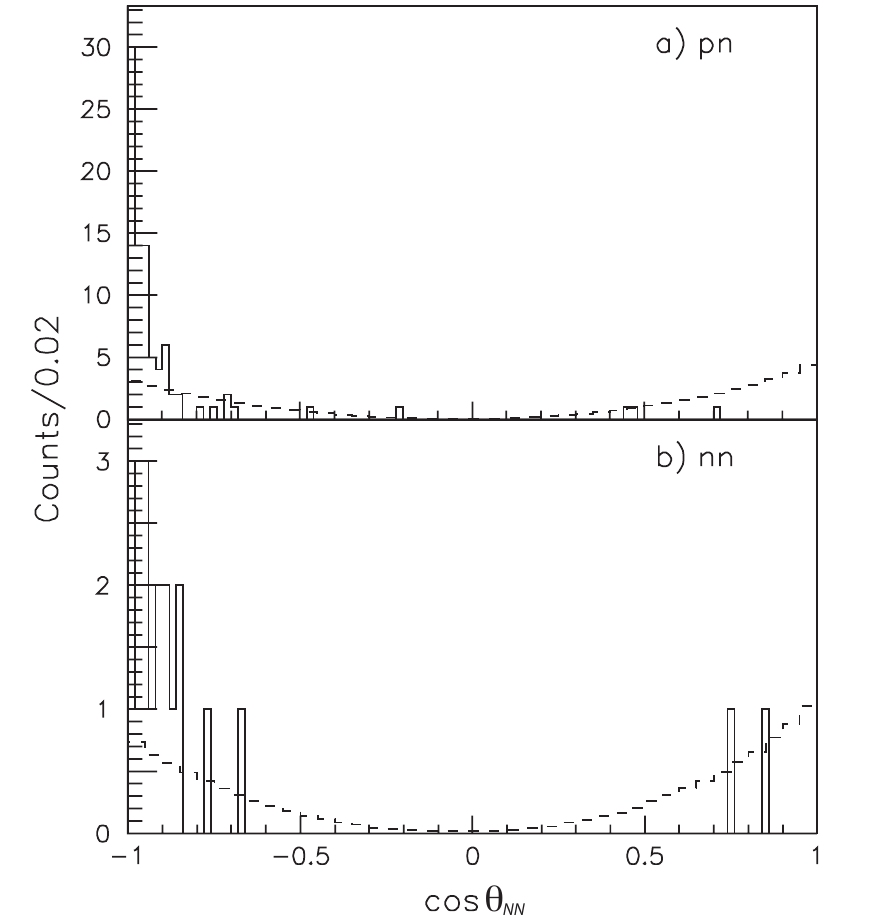}
  \caption[]{ 
The cosine of the nucleon-nucleon separation angle is shown for 
(a) $pn$ and (b) $nn$ coincidence events.
Overlaid on these plots is the relative geometric acceptance for
$N \negthinspace N$ coincidence events normalized to the number of counts
in each spectrum.
\label{fig:coang} }
\end{figure}

\begin{figure}[t]
\includegraphics[width=8.6 cm,clip]{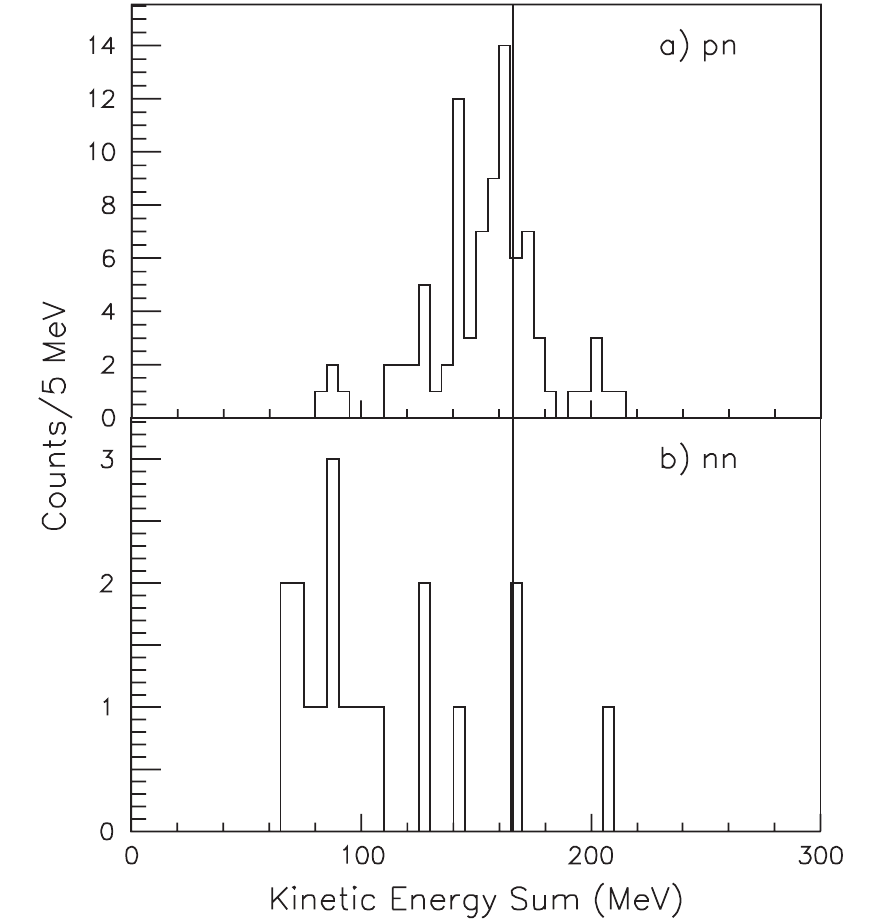}
  \caption[]{ 
The sum of the kinetic energy for the two 
observed nucleons is shown for (a) $pn$ and (b) $nn$ events.
The line is drawn at the expected value of 166~MeV.
\label{fig:coke} }
\end{figure}

The cosine of the separation angle for the two final-state nucleons,
$\cos \theta_{N \negthinspace N}$, is shown in Figures~\ref{fig:coang}(a) 
and \ref{fig:coang}(b) for $pn$ and $nn$ coincidence events, respectively.
If the spectator nucleons are ignored, the two nucleons resulting from
the \LN interaction would be expected to emerge roughly back-to-back
with a separation angle near 180$^\circ$ (from simple momentum conservation).
The observed distribution for $pn$ events is strongly peaked near the value
of 180$^\circ$ with only
about 10\% of the $pn$ events having a separation angle less than 140$^\circ$.
The latter events may be indicative of FSI or \LNN decay contributions.
The number of counts is low in the $nn$ spectrum, but
the back-to-back peak does not seem to be as pronounced for these events.
The distributions for the sum of the kinetic energies of the coincident
nucleons are shown in Fig.~\ref{fig:coke}.
For the \LN interaction,
the total energy available to the two decay nucleons is $\sim$166~MeV 
({\em i.e.}, the difference in the $\Lambda$ and $n$ masses less the binding
energy of the \LfourHe hypernucleus).
The $pn$ kinetic energy distribution of Fig.~\ref{fig:coke}(a) peaks near 
166~MeV as expected.
The observed spread is largely consistent with a spread due to
the Fermi momentum of the baryons within the nucleus coupled with detector
resolution, and
the apparent tail out to lower energies may be the result of FSI or 
\LNN decays.
(The observed spectrum is in qualitative agreement with the results of a
Monte Carlo simulation which includes such effects.)
The $nn$ kinetic energy distribution of Fig.~\ref{fig:coke}(b), 
on the other hand, does not behave
as expected and may be indicative of $nn$ coincidence events originating
from sources other than \Ln decays ({\em i.e.}, some mixture of background neutrons
and neutrons from \Lp or \LNN decays).

The nucleon-stimulated branching fractions were extracted from the observed
$pn$ and $nn$ coincidence events by relating the number of each type of
coincidence event to the branching fractions as:
\begin{eqnarray}
N_{pn} &=& \left( P_{acc}^{(1)} P_{p}^{\Lambda p} + P_{pn}^{\Lambda p} \right)
 B_p g_2 N^{tot}_{\text{HN}}, \\
N_{nn} &=& \left( P_{acc}^{(1)} P_{n}^{\Lambda n} + P_{nn}^{\Lambda n} \right)
 B_n g_2 N^{tot}_{\text{HN}} + N^{bg}_{nn} \label{eq:nnn},
\end{eqnarray}
where $g_2$ is the fraction of a Gaussian distribution contained within
$\pm 2\sigma$ of the mean,
$P_{acc}^{(1)}$ is the probability of detecting exactly one accidental
neutron per event,
$P_{p}^{\Lambda p}$ is the probability of detecting one proton from \Lp decay,
$P_{pn}^{\Lambda p}$ is the probability of detecting one proton and one neutron
from \Lp decay, and
$P_{n}^{\Lambda n}$ and $P_{nn}^{\Lambda n}$ are the probabilities of
detecting one and two neutrons from \Ln decay, respectively.
With the exception of $P_{acc}^{(1)}$, which was extracted from the data,
the above detection probabilities had to be estimated using the Monte Carlo 
simulation described in the previous section.
Final-state interactions were included in the Monte Carlo at the level
determined for the case of FSI only.

The term $N^{bg}_{nn}$ appearing in Eq.~\ref{eq:nnn}
is the expected number of $nn$ background events due to 
accidental neutrons, neutrons from \pim absorption, and
neutrons from proton-stimulated decays.
As with the accidental detection probability above, the detection probabilities
for neutrons from \pim interactions were determined from the data with no
presumptions about the extent of FSI or \LNN decays, while
the detection probabilities for neutrons from \Lp decay were again estimated
using the Monte Carlo with FSI included.
The number of expected background $nn$ coincidence events was found to be
$N_{nn}^{bg} = 18.3 \pm 6.1$ with the largest contribution arising from
neutrons from \pim absorption ($\sim$13 events).
Such a large background is consistent with the observed $nn$ energy-sum 
spectrum of Fig.~\ref{fig:coke}(b).

The resulting values for the branching fractions are listed in 
Table~\ref{tab:svm} along with the results from the analysis of the 
single-particle spectra of the previous section for the case of FSI only.
Also shown is the weighted average of the two determinations.

\begingroup
\squeezetable
\begin{table}[ht]
\caption{\label{tab:svm} Nonmesonic decay rates as determined from the
single-particle kinetic energy spectra and multiple nucleon coincidence
data.
The errors given consist of statistical plus systematic errors, and
upper limits are given at the 95\% CL.
Also shown is the weighted average of the two determinations.}
\begin{ruledtabular}\begin{tabular}{lccc}
 & Single-Particle & Multiple Coincidence & Average \\
\hline
$\Gamma_p/\Gamma_{tot}$ & $0.160 \pm 0.018$ 
& $0.165 \pm 0.028$ & $0.161 \pm 0.015$ \\
$\Gamma_n/\Gamma_{tot}$ & $-0.0089 \pm 0.0135$
& $0.0031 \pm 0.0338$ & $-0.0073 \pm 0.0125$ \\
 & $\leq 0.018$ & $\leq 0.070$ & $\leq 0.018$ \\
$\Gamma_{nm}/\Gamma_{tot}$ & $0.151 \pm 0.017$
& $0.168 \pm 0.042$ & $0.153 \pm 0.016$ \\
$\Gamma_n/\Gamma_p$ & $-0.056 \pm 0.082$
& $0.019 \pm 0.205$ & $-0.046 \pm 0.076$ \\
 & $\leq 0.11$ & $\leq 0.42$ & $\leq 0.11$ \\
\end{tabular}
\end{ruledtabular}\end{table}
\endgroup


In addition to the $pn$ and $nn$ events, a handful of
$pp$ coincidence events were also observed
and were used to set an upper limit on the \Lpp branching fraction.
After the in-beam tracking cuts and $\pm 2\sigma$ excitation energy cut,
a total of 8 $pp$ coincidence events remained.
Although these events may be indicative of a $\Lambda pp$ three-body decay mode,
they could also be $\Lambda p$ decay events for which the 
spectator proton has been detected.
As in the $pn$ and $nn$ cases, the number of $pp$ events of each type may 
be expressed in terms of the \LfourHe branching fractions as:
\begin{eqnarray}
N_{pp}^{\Lambda p} &=& g_2 \varepsilon_q P_{pp}^{\Lambda p} B_p 
 N^{tot}_{\text{HN}} \label{eq:nlp}, \\
N_{pp}^{\Lambda pp} &=& g_2 \varepsilon_q P_{pp}^{\Lambda pp} B_{pp} 
 N^{tot}_{\text{HN}} \label{eq:nlpp},
\end{eqnarray}
where $\varepsilon_q = 0.986$ is the measured efficiency of the charged 
trigger,
$P_{pp}^{\Lambda p}$ and $P_{pp}^{\Lambda pp}$ are the probabilities of 
detecting two protons from \Lp and \Lpp decay, respectively,
and $B_{pp}$ is the \Lpp branching fraction.
The probability of detecting two protons from \Lp decay is negligibly small
when interactions with the spectator nucleons are ignored, but when FSI is
included at the level determined in Sec.~\ref{sec:nmes}, it becomes
significant.
The probabilities $P_{pp}^{\Lambda p}$ and $P_{pp}^{\Lambda pp}$ were
estimated using the Monte Carlo simulation discussed in Sec.~\ref{sec:nmes},
including FSI for the \Lp case.
The expected number of background $pp$ events, $N_{pp}^{\Lambda p}$, was
then calculated from Eq.~\ref{eq:nlp} as $6.40 \pm 0.69$.
Assuming Poisson statistics and 
using the prescription of Feldman and Cousins~\cite{Feldman98}, the upper
limit for the number of observed $pp$ events originating from \Lpp decay
was found as $N_{pp}^{\Lambda pp} \leq 9.29$ (95\% CL for a total number
of 8 observed counts and an expected background of 6 counts).
Using Eq.~\ref{eq:nlpp}, the upper limit for the \Lpp branching fraction
was then found as $B_{pp} \leq 0.026$.

As mentioned above, the contribution from \Lp decay events becomes
negligible when final-state interactions are ignored.
In this case, all 8 observed 
$pp$ events are considered to have originated from $\Lambda p p$ decay, and
the \Lpp branching fraction 
is found to lie in the range $0.008 < B_{pp} < 0.041$ at the 95\% CL.

\subsection{Combined \LfourHe decay rates}

The final values for the \LfourHe branching fractions were determined by 
considering a set of parameters ($B_{\pi^-}$, $B_{\pi^\circ}$, $B_p$, 
$B_n$, $N$) representing the $\pi^-$, $\pi^\circ$, proton- and 
neutron-stimulated branching fractions along with the total number
of \LfourHe hypernuclei.
The optimal values for this set of parameters were found
by minimizing a $\chi^2$ function of the form:
\begin{equation}
\begin{split}
\chi^2 =& \frac{\left(M_{\pi^-} - B_{\pi^-} N\right)^2}{\sigma_{\pi^-}^2}
 +  \frac{\left(M_{\pi^\circ} - B_{\pi^\circ} N\right)^2}{\sigma_{\pi^\circ}^2}
\\
&+  \frac{\left(M_p - B_p N\right)^2}{\sigma_p^2}
 +  \frac{\left(M_n - (B_p + 2 B_n) N\right)^2}{\sigma_n^2} \\
&+  \frac{\left(M_{pn} - B_p N\right)^2}{\sigma_{pn}^2}
 +  \frac{\left(M_{nn} - B_n N\right)^2}{\sigma_{nn}^2} \\
&+  \frac{\left(N_{\text{HN}}^{tot} - N\right)^2}{\sigma_N^2},
\end{split}
\end{equation}
where
the values $M_a$ are the acceptance-corrected numbers of event type
$a$ with associated error $\sigma_a$.
Additionally,
the parameters representing the branching fractions were subject to the
constraint:
\begin{equation}
B_p + B_n + B_{\pi^-} + B_{\pi^\circ} = 1,
\end{equation}
and any possible contributions from three-body \LNN decay, which were previously 
seen to be small, were neglected.
The minimization was carried out using the ROOT MINUIT minimization routine,
and the results
are listed in 
Table~\ref{tab:res1} (in units of $\Gamma_{tot}$).

\begin{table}[t]
\caption{\label{tab:res1} The \LfourHe decay rates as determined
by the present analysis.
The errors quoted include contributions from systematic error, and
upper limits are given at the 95\% CL.
}
\begin{ruledtabular}\begin{tabular}{lclc}
\multicolumn{2}{c}{Mesonic rates} & \multicolumn{2}{c}{Nonmesonic rates} \\
\hline
$\Gamma_{\pi^\circ}/\Gamma_{tot}$ & $0.564 \pm 0.036$ &
$\Gamma_p/\Gamma_{tot}$ & $0.169 \pm 0.019$ \\
$\Gamma_{\pi^-}/\Gamma_{tot}$ & $0.270 \pm 0.024$ & 
$\Gamma_n/\Gamma_{tot}$ & $-0.0032 \pm 0.0183$ \\
 & & & $\leq 0.032$ \\
$\frac{\Gamma_{\pi^\circ} + \Gamma_{\pi^-}}{\Gamma_{tot}}$ & $0.835 \pm 0.021$ & 
$\Gamma_{nm}/\Gamma_{tot}$ & $0.165 \pm 0.021$ \\
$\Gamma_{\pi^\circ}/\Gamma_{\pi^-}$ & $2.09 \pm 0.29$ &
$\Gamma_n/\Gamma_p$ & $-0.019 \pm 0.108$ \\
 & & & $\leq 0.19$ \\
\end{tabular}
\end{ruledtabular}\end{table}

\section{Discussion}
\label{sec:conc}

The \LfourHe decay rates determined in the present work 
are listed in 
Table~\ref{tab:res} (in units of $\Gamma_\Lambda$).
Also listed are the results of another 
measurement of \LfourHe decay performed
by Outa {\em et al}~\cite{Outa98} along with the results of some 
earlier bubble chamber experiments~\cite{Reinhard92,Block64,Fetkovich72}.
The values taken from reference~\cite{Reinhard92} are the result of a 
reanalysis of various older results performed by Schumacher.
The decay rates found in this work are seen to be in good agreement with past
measurements.

The results reported for the experiment of Outa {\em et al} were published
in 1998, some 8 years after the running of our experiment.
Their experiment created \LfourHe by the 
$^4\rm{He}(K^-_{stopped},\pi^-)$ reaction and detected the
$\pi^\circ$, $\pi^-$, and protons from the subsequent weak decay.
Unlike our experiment, their apparatus was not able to detect neutrons, 
requiring them to determine the neutron-stimulated rate by subtraction.
Our results for the neutron-stimulated rate and neutron to proton ratio, 
$\Gamma_n/\Gamma_\Lambda \leq 0.035$ and $\Gamma_n/\Gamma_p \leq 0.19$,
provide stricter upper limits as compared to 
$\Gamma_n/\Gamma_\Lambda \leq 0.09$ and $\Gamma_n/\Gamma_p \leq 0.60$ 
for Outa {\em et al} (all at the 95\% CL).
The upper limits for Outa {\em et al} given above were derived from the 
results listed in Table~\ref{tab:res} using the same method as for our results.
There is also a slight difference in the mesonic decay rates, where our 
$\Gamma_{\pi^\circ}/\Gamma_{\pi^-}$ ratio is seen to be in better agreement
with the earlier bubble chamber results.

Selected theoretical investigations
of the \LfourHe nonmesonic partial decay rates are shown in
Table~\ref{tab:theo}.  
These OPE (one pion exchange) models incorporate a quark-based model for
short range interactions and are currently considered to be the most successful
of the OPE-based models.
The results of Jun~\cite{Jun01} and Inoue {\em et al}~\cite{Inoue98} are shown 
for OPE only and OPE + quark interaction mechanism.
The work of Jun utilizes a phenomenological 4-baryon point interaction model
to describe the short range quark interactions, while Inoue {\em et al}~uses a 
direct quark (DQ) interaction based on the effective weak Hamiltonian of
Gilman and Wise~\cite{Gilman79}.
The calculation of Sasaki {\em et al}~\cite{Sasaki05} employs the direct quark
mechanism of Inoue {\em et al}~coupled with a $\pi + K + \sigma$ exchange 
potential.
Unlike previous models, this model is able to reproduce the value of the 
proton asymmetry observed in polarized \LfiveHe nonmesonic decay.
All of the above models adequately reproduce the observed \LfourHe nonmesonic
decay rates.

In regards to the \delIeq rule, it is interesting to note
that the direct quark models predict a significant 
contribution from \delIeqthree transitions arising from the effective
weak Hamiltonian used, 
while models based on the 
phenomenological 4-baryon point interaction are able to reproduce the
experimental results with little or no \delIeqthree component.
All of these models enforce the \delIeq rule for the meson-exchange component.

\begin{table}
\caption{\label{tab:res} Measured properties of the \LfourHe hypernucleus.
Listed here are the results of the present analysis along with
the results of another recent experiment on \LfourHe decay~\cite{Outa98} 
and some older results from bubble chamber
experiments~\cite{Block64,Fetkovich72,Keyes76}.
The errors quoted for this analysis include contributions from systematic
error.}
\begin{ruledtabular}\begin{tabular}{llll}
 & This work & Outa {\em et al}~\cite{Outa98} 
& Earlier results \\
\hline
$\Gamma_{tot}/\Gamma_\Lambda$ & $1.07 \pm 0.11$ & $1.03^{+0.12}_{-0.10}$ &
$1.15 \pm 0.48$~\cite{Reinhard92} \\
$\Gamma_{\pi^\circ}/\Gamma_\Lambda$ & $0.604 \pm 0.073$ 
 & $0.53 \pm 0.07$ & \\
$\Gamma_{\pi^-}/\Gamma_\Lambda$ & $0.289 \pm 0.039$ 
 & $0.33 \pm 0.05$ & \\
$\Gamma_{\pi^\circ}/\Gamma_{\pi^-}$ & $2.09 \pm 0.29$ 
 & $1.59 \pm 0.20$ & $2.49 \pm 0.34$~\cite{Block64} \\
 & & & $2.20 \pm 0.39$~\cite{Fetkovich72} \\
$\Gamma_{\pi^+}/\Gamma_{\pi^-}$ & & & $0.043 \pm 0.017$~\cite{Keyes76} \\
$\Gamma_p/\Gamma_\Lambda$ & $0.180 \pm 0.028$ 
 & $0.16 \pm 0.02$ & \\
$\Gamma_n/\Gamma_\Lambda$ & $\leq 0.035$
 & $0.01^{+0.04}_{-0.01}$ & \\
$\Gamma_{nm}/\Gamma_\Lambda$ & $0.177 \pm 0.029$
 & $0.17 \pm 0.05$ & \\
$\Gamma_{nm}/\Gamma_{\pi^-}$ & $0.61 \pm 0.10$
 & $0.51 \pm 0.16$ & $0.56 \pm 0.09$~\cite{Reinhard92} \\
$\Gamma_n/\Gamma_p$ & $\leq 0.19$
 & $0.06^{+0.28}_{-0.06}$ & $0.40 \pm 0.15$~\cite{Block64} \\
 & & & $0.29 \pm 0.13$~\cite{Fetkovich72} \\
\end{tabular}
\end{ruledtabular}\end{table}

\begin{table}
\caption{\label{tab:theo} Results of selected theoretical investigations
of the \LfourHe partial decay rates.  
The partial rates listed here are given in units of the free $\Lambda$ 
decay width $\Gamma_\Lambda$.}
\begin{ruledtabular}\begin{tabular}{lccccc}
 & \multicolumn{2}{c}{Jun~\cite{Jun01}} 
 & \multicolumn{2}{c}{Inoue~\cite{Inoue98}}
 & Sasaki~\cite{Sasaki05} \\
 & OPE & OPE+4BPI & OPE & OPE+DQ & DQ+ \\
\hline
$\Gamma_n/\Gamma_\Lambda$ & 0.0008 & 0.0373 & 0.009 & 0.038 & 0.015 \\
$\Gamma_p/\Gamma_\Lambda$ & 0.1478 & 0.1602 & 0.145 & 0.214 & 0.169 \\
$\Gamma_{nm}/\Gamma_\Lambda$ & 0.1486 & 0.1975 & 0.154 & 0.253 & 0.184 \\
$\Gamma_n/\Gamma_p$ & 0.005 & 0.23 & 0.061 & 0.178 & 0.091 \\
\end{tabular}
\end{ruledtabular}\end{table}

\subsection{FSI, three-body \LNN decays, and interference effects}

Using the simple model for nonmesonic decay introduced in 
Sec.~\ref{sec:nmes}, the 
contributions of final-state interactions and three-body \LNN decays were
studied utilizing the single-particle proton and neutron kinetic energy
spectra, and upper limits for each case were determined as:
\begin{equation}
\frac{\Gamma^{\text{FSI}}_{nm}}{\Gamma_{tot}} \leq 0.11,
\qquad \text{and} \qquad
\frac{\Gamma_{mb}}{\Gamma_{tot}} \leq 0.097,
\end{equation}
for final-state interactions and \LNN decays, respectively (95\% CL).
Within the model used here, the effects of FSI and \LNN decays on
pertinent quantities were seen to be small and, thus, have little effect on 
the final interpretation of the data.


In the analysis of the nonmesonic decay rates, the various decay
modes have been assumed to add incoherently, neglecting possible quantum
interference effects.
It has been suggested~\cite{Garbarino04,Oka04} that such interference 
effects may be the origin of the so-called \Gnpr puzzle, and
Garbarino {\em et al}~\cite{Garbarino04} point out that any interference effects
would be less pronounced for multiple coincidence measurements, providing
a cleaner extraction of the nonmesonic decay rates.
For the case of \LfourHe presented here, the results of the two 
determinations (listed in Table~\ref{tab:svm}) are consistent, suggesting 
that interference effects are not significant at this level of statistics.
A detailed study of the angular correlations in the multiple nucleon
coincidence data 
might shed more light on the contributions of interference effects as well as
FSI and \LNN decays, but
the number of coincidence events in the present data set is too limited 
to provide anything conclusive.

\subsection{Phenomenological analysis of $s$-shell hypernuclei}

As mentioned in the introduction, 
measurements of the nonmesonic decay rates for the $s$-shell hypernucleus 
\LfourH are currently quite limited.
The most recent results from Outa {\em et al}~\cite{Outa98} 
give a value for the total nonmesonic rate of 
$\Gamma_{nm}/\Gamma_\Lambda(^4_\Lambda\text{H}) = 0.17 \pm 0.11$, but
no measurements of the proton- and neutron-stimulated partial rates
have been made.
Using the present results for $^4_\Lambda\text{He}$, these partial rates 
can be calculated
in the context of the previously described phenomenological model
of Block and Dalitz.

From Eq.~\ref{eq:hp}, the upper limit of 
$\Gamma_n/\Gamma_\Lambda(^4_\Lambda\text{He}) \leq 0.035$ determined
for the current measurement implies an upper limit of 
$\Gamma_p/\Gamma_\Lambda(^4_\Lambda\text{H}) \leq 0.017$ for the \LfourH 
proton-stimulated rate under the assumption of pure 
$\Delta I = \frac{1}{2}$ transitions.
A measurement of the \LfourH
proton-stimulated rate greater than this value would indicate a 
violation of the \delIeq rule for the \LN weak interaction.
The \LfourH neutron-stimulated decay rate was found from Eq.~\ref{eq:hn} as
$\Gamma_n/\Gamma_\Lambda(^4_\Lambda\text{H}) = 0.081 \pm 0.023$.
This calculation utilizes the result of 
Outa {\em et al}~for the ratio of the neutron- to proton-stimulated rates for
$^5_\Lambda$He, $\Gamma_n/\Gamma_p(^5_\Lambda\text{He}) 
 = 0.45 \pm 0.11 \pm 0.03$~\cite{Outa05}, but makes no assumptions regarding
the \delIeq rule.
Using this result, 
an alternate determination of the \LfourH proton-stimulated rate can be made 
from the result of 
Outa {\em et al}~for the total nonmesonic decay width of $^4_\Lambda\text{H}$:
\begin{equation}
\Gamma_p(^4_\Lambda\text{H}) = \Gamma_{nm}(^4_\Lambda\text{H})
 - \Gamma_n(^4_\Lambda\text{H}) = (0.09 \pm 0.11)\Gamma_\Lambda,
\end{equation}
which, again, is independent of any assumptions about the isospin nature of
the \LN interaction.
Because of the large error, this result is consistent 
with the upper limit found above.
Thus, at the present level of statistics for the \LfourH measurement,
nothing conclusive can be said about the \delIeq rule for the \LN weak
interaction.

The values of the \LfourH nonmesonic decay rates derived from the 
above discussion represent the most current knowledge (although indirect) of
the \LfourH proton- and neutron-stimulated partial rates.
The upper limit determined for the \LfourH proton-stimulated decay rate
suggests a direct measurement of this decay mode as an easy test of the 
\delIeq rule for nucleon-stimulated decays ({\em i.e.}, a measurement in excess of
this upper limit would show a clear violation of the \delIeq rule for the
\LN weak interaction).
Measurements of the \LfourH nonmesonic decay rates along with more precise
determinations of the \LfourHe and \LfiveHe nonmesonic decay rates would
help to further determine the isospin structure of the \LN weak interaction.
Multiple nucleon coincidence studies 
can potentially
provide the cleanest extraction of the nucleon-stimulated partial rates
and would be a worthwhile avenue of future research.

\begin{acknowledgments}

This work was partially supported by U.S. DOE grant nos. 
DE-FG02-87ER40315, 
DE-AC02-98CH10886, 
and DE-FG03-94ER40821.

\end{acknowledgments}

\end{document}